\documentclass[aoas,preprint]{imsart}

\RequirePackage[colorlinks,citecolor=blue,urlcolor=blue]{hyperref}

\startlocaldefs
\usepackage{amsmath}
\usepackage{graphicx}
\usepackage{enumerate}
\usepackage{natbib}
\usepackage{url} 

\usepackage{graphicx}
\usepackage{natbib}
\usepackage{amsfonts,amssymb,amsbsy,amsthm,mathtools}
\usepackage{bbm,bm}
\usepackage{color}
\usepackage{subfig,float}
\usepackage{booktabs}
\usepackage{paralist}
\allowdisplaybreaks
\numberwithin{equation}{section}

\newtheorem{thm}{Theorem}[section]

\newtheorem{prop}[thm]{Proposition}
\theoremstyle{definition}

\newcommand{\RR}{\mathbb{R}}
\newcommand{\PP}{\mathbb{P}}
\newcommand{\EE}{\mathbb{E}}
\newcommand{\vc}[1]{\bm{#1}}

\newcommand{\emp}{\textnormal{emp}}
\endlocaldefs

\begin{document}

\begin{frontmatter}

\title{Climate extreme event attribution using 
 multivariate peaks-over-thresholds modeling 
 and counterfactual theory}
\runtitle{Climate extreme event attribution}

\begin{aug}
\author{\fnms{Anna} \snm{Kiriliouk}\thanksref{t1,m1}\ead[label=e1]{anna.kiriliouk@unamur.be}}
\and
\author{\fnms{Philippe} \snm{Naveau}\thanksref{m2}\ead[label=e2]{naveau@lsce.ipsl.fr}}

\affiliation{Universit\'e de Namur\thanksmark{m1} and CNRS (Gif-sur-Yvette) \thanksmark{m2}}
\thankstext{t1}{Part of this work was supported by the French national program FRAISE-LEFE/INSU, MELODY-ANR, 
Eupheme, FUSIMET (PEPS I3A) and the DAMOCLES-COST-ACTION on compound events.}
\runauthor{A. Kiriliouk and P. Naveau}
\end{aug}

\begin{abstract}
Numerical climate models  are complex and combine a large number of physical processes. 
They are key tools in quantifying the relative contribution of potential anthropogenic causes (e.g., the current increase in greenhouse gases) on high impact atmospheric variables like heavy rainfall. These so-called climate extreme event attribution problems are particularly challenging in a multivariate context, that is, when the atmospheric variables are measured on a possibly high-dimensional grid.

In this paper, we leverage two statistical theories to assess causality in the context of multivariate extreme event attribution. As we consider an event to be extreme when at least one of the components of the vector of interest is large,   
extreme-value theory justifies, in an asymptotical sense, a multivariate generalized Pareto distribution to model joint extremes.
Under this class of distributions, we derive and study probabilities of necessary and sufficient causation as defined by the counterfactual theory of Pearl. To increase causal evidence, we propose a dimension reduction strategy based on the optimal linear projection that maximizes such causation probabilities.  Our approach is tested on simulated examples and applied to weekly winter maxima precipitation outputs of the French CNRM from the recent CMIP6 experiment.  
\end{abstract}



\end{frontmatter}

\section{Introduction}
Quantifying human influence on climate change and identifying potential causes of climate  
extremes is often referred  to as extreme event attribution (EEA), which falls within the research field of 
 detection and attribution (D\&A)  \cite[see, e.g. the report of ][]{NAP16,Chen18}.  
 In such studies, the main inferential objective is estimation of extreme quantiles (also called return levels), well-used in  finance, hydrology and other fields of risk analysis \citep[e.g.][]{Embrechts97}. 
In EEA, 
return levels are computed under two scenarios that differ according to the causal link of interest, often increases in greenhouse gases (GHG) concentrations \cite[see, e.g.][]{Angelil17,stott16-review,fischer2015}. 
Typically, such an approach compares probabilities under a factual scenario of conditions that occurred around the time of the event against probabilities under a counterfactual scenario in which anthropogenic emissions never occurred. More specifically, one compares the probability of an extreme event in the factual world, denoted $p_1$, to the probability of an extreme event $p_0$ in a counterfactual world, i.e., a world that might have been if no anthropogenic forcing would have been present.
The definition of the so-called extreme event is by itself a non-trivial task and depends on the application at hand. A common choice is to take some climatological index exceeding a high threshold. In their seminal paper,  \cite{Stott04} studied mean June--August temperatures in Europe in order to quantify by how much human activities may have increased the risk of European heatwaves. In this example, a one dimensional sample mean, say $X$, summarized a complex random field that varied in time and space. 
The set $\{ X >v \}$ where $v= 1.6 $ Kelvin was chosen to resemble the 2003 mean European summer anomaly temperatures. 
The probabilities $p_0$ and $p_1$ were then inferred from numerical counterfactual and factual runs respectively, using
nonparametric inference techniques \citep[for bootstrap counting methods in EEA, see][]{Paciorek18} 
and univariate extreme-value theory (EVT); see, e.g. \citet{Coles01} for an introduction. 

In other environmental research areas, complex multivariate 
EVT models are commonly used
\cite[see, e.g.][]{Davison15,Cooley17,Fondeville18,Engelke19,Reich13,Shaby12}.
Bayesian hierarchical modeling  \citep[see, e.g.][]{Hammerling17,Katzfuss17}  also offers a flexible way to insert different layers of complexity present in climate D\&A  problems  (internal natural variability, 
numerical model uncertainty, observational errors, sampling uncertainty in space and time,  etc.). 
Despite these advances, the EEA domain remains a fairly untouched territory in terms of multivariate EVT. Even recent papers like  \cite{Kew19,Luu18,Otto18} and \cite{King17} are based on univariate EVT only.

Still, in climatological studies, there is no reason to assume that the dependence structure remains identical over space. 
For example, heavy rainfall in convective prone regions will spatially differ from regions driven by frontal storms. 
Examples like the one shown  in Table 1  below underscore  that the practitioner choice  of  the region under study  and the risk measure used can lead to very different return periods. 
In addition, the essence of EEA studies is to detect changes between return periods in factual and counterfactual worlds  \cite[see, e.g.,][ for a recent review on the statistical aspects of this problem]{naveau20}.
If the spatial dependence structure differs between the factual and counterfactual worlds (e.g., a region could be more prone to convective storms in its factual version), it has an influence on the return level. 
Hence, past studies based on univariate EVT measures may have  overlooked   some causal signals by not taking into account the underlying multivariate structure.

Our first objective is to investigate how multivariate EVT could be used for EEA. As extreme events in D\&A  are mostly expressed in terms of threshold exceedances, like $\{ X >v \}$ in \cite{Stott04}, this naturally leads to the question of how to integrate 
the multivariate generalized Pareto distribution (GPD)  introduced by \citet{tajvidi1996} into the EEA framework. 
This  distribution has been tailored to represent extremal behaviors when at least one of the components of the vector of interest is large.
The probabilistic properties of the multivariate GPD have been well studied by, among others, \citet{beirlant2004,rootzen2006,falk2008,ferreira2014,
rootzen2018} and \citet{rootzen2018nr2}, while statistical modeling is more recent \citep{huser2015,kiriliouk2018}.
 
In most univariate EEA studies \cite[see][and references therein]{stott16-review}, two types of probability ratios are considered: the Risk Ratio $\frac{p_1}{p_0}$ and the so-called Fraction of Attributable Risk (FAR), defined by 
$$
\mbox{FAR} = 1- \frac{p_0}{p_1}, 
$$
where $p_0  = \PP(X >v)$ corresponds to the probability of exceeding the threshold $v$ in the  counterfactual world, while $p_1$ represents the same quantity in the factual world. 
Using the counterfactual theory of \citet{pearl2000}, the $\mbox{FAR}$ corresponds to the probability of necessary causation, i.e., anthropogenic forcings are necessary for the extreme event to occur, but might not be sufficient.
Within the Gaussian set-up, \citet{hannart2016} and \citet{hannart18}  highlighted  the link between causality theory and event attribution studies. The second objective of our work is to explain how Pearl's counterfactual theory can be applied within a multivariate GPD framework, and to identify conditions that maximize the probability of causality, a fundamental feature in any  EEA analysis. 

The rest of the paper is structured as follows.
Section~\ref{sec:background} summarizes the relevant background of EEA and the multivariate GPD. Section~\ref{sec:univariate} discusses the behaviour of univariate probabilities of causation as a function of the threshold $v$.  
In Section~\ref{sec:multivariate}, we make the link between multivariate GPDs and causality by maximizing necessary causation for any linear projection and we discuss the inference strategy. Finally, in Section~\ref{sec:appli}, the proposed methods are applied to weekly winter maxima of precipitation outputs from the  French CNRM model that are part of  the recent CMIP6 experiment. A discussion and outlook is given in Section~\ref{sec:discussion}. Technical details are deferred to the supplementary material \citep{KirilioukNaveau20}.

\section{Background}\label{sec:background}

\subsection{Climate event attribution and counterfactual theory}\label{sec:introclimate}
The question of attribution in EEA is inherently rooted in causality assessment. \citet{pearl2000} proposed a framework to connect the probabilities $p_0$ and $p_1$ to different forms of causality. If $E$ denotes an event (e.g. the 2003 European heatwave) and $C$ its potential cause (e.g. the increase of GHG emissions), three distinct forms of causality are of interest: 
 \begin{enumerate}
\item probability of necessary causation (PN): $C$ is required for $E$ but other factors might be required as well;
\item probability of sufficient causation (PS): $C$ always triggers $E$ but $E$ might occur without $C$;
\item probability of necessary and sufficient causation (PNS): both of the above hold.
\end{enumerate}
\citet{hannart2016} recalled the mathematical definition of these probabilities,
\begin{align*}
\textnormal{PN} & = \PP[\overline{E} \mid do(\overline{C}), C, E], \quad \textnormal{PS} = \PP[E \mid do(C), \overline{C}, \overline{E}], \\
\textnormal{PNS}  & = \PP[E \mid do(C), \overline{E} \mid do(\overline{C})],
\end{align*}
where $do(C)$ means that the cause is interventionnaly forced to occur.
For climate EEA, the cause $C$ can be defined as the presence of anthropogenic forcings. In this setting,
\citet{hannart2016} exploited that $E$ is monotonous with respect to $C$ (the event is more likely when $C$ is present)
and $C$ is exogenous (i.e., it is not influenced by any other observed variables).  
The causation probabilities then simplify to
\begin{align}\label{eq:probas}
\textnormal{PN}  & = \max \left( 1 - \frac{p_0}{p_1} , 0\right), \quad
\textnormal{PS}  = \max \left( 1 - \frac{1 - p_1}{1-p_0}, 0 \right),  \\
\textnormal{PNS} & = \max \left( p_1 - p_0, 0 \right), \notag 
\end{align}
where $p_0 = \PP [ E \mid \overline{C}]$ corresponds to the probability of $E$ in the counterfactual world and $p_1 = \PP[E \mid C]$ to the probability of the same event in the factual world. If $p_0<p_1$, the PN coincides with the FAR used by \cite{Stott04}. 
In the remainder of this paper, we will use the notation PN (instead of FAR) to highlight its causal interpretation. 
By construction, one has PNS $\leq$ $\min($PN,PS$)$ and hence it is worth noticing that a low PNS does not imply the absence of a causal relationship.

A fundamental step in any causality analysis is the definition of the event $E$.
In  EEA, the event $E$  is classically defined as   the occurrence  of  some climatological  index (e.g.\ a spatial average over a given region) exceeding a high threshold $v$, 
$$
E = \{ \vc{w}^T \vc{X} > v \},
$$ 
where  $\vc{X}=(X_1, \dots, X_d)^T$ is a random vector defined on $d$ grid points and  $ \vc{w} = (w_1,\ldots,w_d)^T$  are non-negative weights. 
Let $\vc{X}^{(0)}$ and $ \vc{X}^{(1)}$ denote the climatological vector $\vc{X}$ in the counterfactual and the factual world respectively. 
Hence,  the probabilities $p_i$ in (\ref{eq:probas}) become
\begin{equation*}
p_i = \PP [\vc{w}^T \vc{X}^{(i)} > v] , \quad i  \in \{ 0,1\}.
\end{equation*}
Generally, $\vc{w}^T \vc{X}^{(i)}$ is modelled as a univariate random variable \cite[see, e.g.][]{Kew19,Luu18,Otto18,King17} 
and the dependence among  the components of $\vc{X}^{(i)}$ is ignored.
One objective of this paper is to explore how  the multivariate dependence among  the $\vc{X}^{(i)}$'s 
affects the values $p_i$, and consequently the causal evidence expressed with PN, PS or PNS, especially if the weights are well chosen. 
In the next section, we present a simple example to gain 
some intuition on the difference in return levels between univariate and multivariate modeling. 

{
\subsection{Impact of multivariate extremal modeling on return periods}
%
 In this section, we illustrate how the dependence structure of $\vc{X}$ impacts the return periods of $\vc{w}^T \vc{X}$. 
 For clarity's sake,  
 we assume that the margins $X_i$ of $\vc{X}$ follow unit exponential distributions
 (in line with the multivariate EVT model of Section~\ref{sec:introGPD}).
 Hence, the associated return level at each grid point is simply $u_T = \log T$  for a given return period $T$, i.e.,  $\PP(X_i>u_T)=1/T$ for $i = 1,\ldots,d$.
If the $X_i$'s exhibit complete dependence, i.e., $X_i = X_j$ with probability 1 for all $i,j$, the return time of $\vc{w}^T \vc{X}$ is identical to that of $X_i$, and is simply equal to $T$.
In the opposite case, the margins of $\vc{X}$ are independent and by construction, $\vc{w}^T \vc{X}$ has unit mean and variance $w^2_1+\dots+w^2_d$. If $w_i=1/d$ for all $i$, 
the return time of $\vc{w}^T \vc{X}$ associated with $u_T $ corresponds to the quantile of a gamma distribution. 
The first and third rows in Table \ref{TableXbar} show that the return periods of the event $\{ w_1 X_1 +  w_2 X_2   > u_T\}$ are much larger than the ones obtained in the complete dependence case. 
This effect increases as the time period increases and/or  the dimension $d$ increases (tables available upon request). Hence, \emph{given the same univariate variable $\vc{w}^T \vc{X}$ and the same return level $u_T = \log T$, the degree of dependence in the original data greatly influences
the return period of the same event}. 
As the field of EEA is rooted in the computation of  events like  $\{ \vc{w}^T \vc{X} > u_T\}$
for large $T$, modeling of multivariate extremal dependence becomes paramount. In the last decades, Rootz\'en and his colleagues  have proposed models and inferential schemes for high return levels that take this dependence into account.
The row ``Intermediate dependence (MGPD)" in  Table \ref{TableXbar} concerns such a multivariate model (to be detailed in Section \ref{sec:introGPD}). 

\begin{table}[ht]
\begin{center}
 \begin{tabular}{llll}
\toprule
& \multicolumn{3}{c}{$w_1 = w_2 = 0.5$} \\
\cmidrule(r){2-4}
& $T = 10$ & $T = 50$ & $T = 100$ \\
\midrule
Complete dependence & 10 & 50 & 100 \\
Intermediate dependence (MGPD) & 19 & 96 & 191 \\
Independence & 18 & 283 & 979 \\
\midrule
& \multicolumn{3}{c}{$w_1 = 0.2, \, w_2 = 0.8$} \\
\cmidrule(r){2-4}
& $T = 10$ & $T = 50$ & $T = 100$ \\
\midrule
Complete dependence & 10 & 50 & 100 \\
Intermediate dependence (MGPD) & 18 & 88 & 175 \\
Independence & 13 & 100 & 237 \\
\bottomrule
\end{tabular}
\end{center}
\caption{Return periods $T$ in years of the event $\{ w_1 X_1 +  w_2 X_2   > u_T\}$ with   $u_T = \log T$ for \\ $T \in$ $ \{10, 50, 100\}$ for $w_1 = w_2 = 0.5$ (top) and $w_1 = 0.2, w_2 = 0.8$ (bottom). \\ The marginal distributions of $X_1$ and $X_2$ are unit exponential satisfying  $P(X_1>u_T)=P(X_2>u_T)=1/T$. The multivariate GPD has tail dependence coefficient $\chi = 0.5$, see Section~\ref{sec:introGPD}.}
\label{TableXbar}
\end{table}

Comparing the upper part of Table \ref{TableXbar} with the lower part, we see that the choice of the weights also plays a non-negligible role in the return times values. 
In EEA, one wishes to contrast the factual and counterfactual worlds. 
Table \ref{TableXbar} shows that differences between $p_0 = \PP [\vc{w}^T \vc{X}^{(0)} > v] $ and $p_1 = \PP [\vc{w}^T \vc{X}^{(1)} > v] $ not only stem from the spatial dependences structures of $\vc{X}^{(0)}$ and $\vc{X}^{(1)}$, but also from the weight selection.

\subsection{The multivariate generalized Pareto distribution}\label{sec:introGPD}
When $\vc{X} = X \in \RR$, univariate peaks-over-thresholds approaches 
\citep{davison1990} consist of choosing a large threshold $u$ and fitting $\left[ X -u  \mid X > u  \right]$  to a univariate GPD. Hence, regardless of the underlying distribution of the climatological index $X$, the GPD can be used to model causation probabilities of the event $\{X > v \}$ as long as $v > u$.
Similarly, a multivariate GPD approximates the tail behavior of  $\left[ \vc{X} - \vc{u} \mid \vc{X} \nleq \vc{u}  \right]$, where $ \vc{X} \nleq \vc{u} $ means that at least one component of $\vc{X}$ exceeds the corresponding component of the threshold $\vc{u} \in \mathbb{R}^d$. In the following, we will see how it can be used to model causation probabilities of the event $\{ \vc{w}^T \vc{X} > v\} $ for $v > \vc{w}^T \vc{u}$.

From a mathematical point of view, multivariate GPD vectors can be viewed as the limiting solution of any linearly rescaled multivariate vector given that at least one component is large. 
This asymptotic result can be interpreted as a multivariate version of the Pickands--Balkema--de Haan theorem  \citep{pickands1975,balkema1974}. 
For the sake of clarity and concision, we will only recall how multivariate GPDs can be simulated and how basic principles are derived from this stochastic definition. 
The reader interested by theoretical aspects of multivariate GPDs is referred to \cite{rootzen2006}. 

The basic building block to construct standardized multivariate GPD vectors \citep{rootzen2018} is the  stochastic representation
\begin{equation}\label{eq:Trep}
\vc{Z}^* \overset{\mathrm{d}}{=}  E + \vc{T} - \max_{1 \le j \le d} T_j,
\end{equation}
where $E$ is a unit exponential random variable and $\vc{T}=(T_1,\dots, T_d)^T$ represents any $d$-dimensional random vector independent of $E$. 
One can check that each positive conditional margin has a unit exponential survival function,
$$
 \PP [Z^*_j > z \mid Z^*_j > 0] =  \exp( -z ), \quad \mbox{ for any $z>0$ and $j \in \{1,\ldots,d \}$.}
$$
Model (\ref{eq:Trep}) can be generalized by setting, for $\vc{\sigma}>\vc{0}$ and $\vc{\gamma}  \in \RR^d$,
\begin{align}
\vc{Z} \overset{\mathrm{d}}{=} \frac{\vc{\sigma}}{\vc{\gamma}} \left( \exp\left( \vc{\gamma} \vc{Z}^* \right) -1 \right),   
\label{eq:gpconstr}
\end{align}
where operations like $\frac{\vc{\sigma}}{\vc{\gamma}}$ have to be understood componentwise. 
We then denote $\vc{Z} \sim \textnormal{MGPD}(\vc{T},\vc{\sigma},\vc{\gamma})$. Equation (\ref{eq:gpconstr}) implies
\begin{align*}
 \PP [Z_j > z \mid Z_j > 0] = \overline{H}( z ; \sigma_j, \gamma_j), \quad \mbox{ for any $z>0$ and $j \in \{1,\ldots,d \}$,}
\end{align*}
where $\overline{H}( z ; \sigma, \gamma)=(1+ \gamma z /\sigma)^{-1/\gamma}_+$ denotes the survival function of the univariate GPD with scale parameter $\sigma > 0 $ and shape parameter $\gamma \in \mathbb{R}$.   
Hence, the conditional margins $\left[ Z_j \mid Z_j > 0 \right] $ follow univariate GPDs\footnote{although the random variables $Z_1,\ldots,Z_d$ may not follow GPDs themselves. 
}.

The random ``generator" $\vc{T}$ in (\ref{eq:Trep}) drives the extremal dependence of $\vc{Z}$, often summarized by the \emph{tail dependence coefficient} $\chi \in [0,1]$ (for $d = 2$). If $F_1$, $F_2$ denote the unconditional marginal distribution functions of $Z_1, Z_2$, then $\chi$
measures the probability of $F_1 (Z_1)$ being large given that $F_2 (Z_2)$ is large as the threshold increases, 
\begin{align*}
\chi & = \lim_{q \uparrow 1} \PP \left[ F_1 (Z_1) > q \mid F_2 (Z_2) > q \right] \\
& = \EE \left[ \min \left( \frac{e^{T_1 - \max(T_1,T_2)}}{\EE \left[ e^{T_1 - \max(T_1,T_2)} \right]}, \frac{e^{T_2 - \max(T_1,T_2)}}{\EE \left[ e^{T_2 - \max(T_1,T_2)} \right]} \right) \right].
\end{align*}
A large value of $\chi$ corresponds to strong extremal dependence between $Z_1$ and $Z_2$, whereas $\chi = 0$ corresponds to tail independence. For more details on $\chi$ in the context of multivariate GPDs, see the supplementary material in \citet{kiriliouk2018}. As an example, Figure \ref{fig:gauss} displays 500 bivariate random draws from a multivariate GPD model where $\vc{T}$ is zero-mean bivariate Gaussian with unit covariance matrix $I_2$, corresponding to $\chi = 0.6$.

\begin{figure}[ht!]
\centering
\subfloat{\includegraphics[scale=.35]{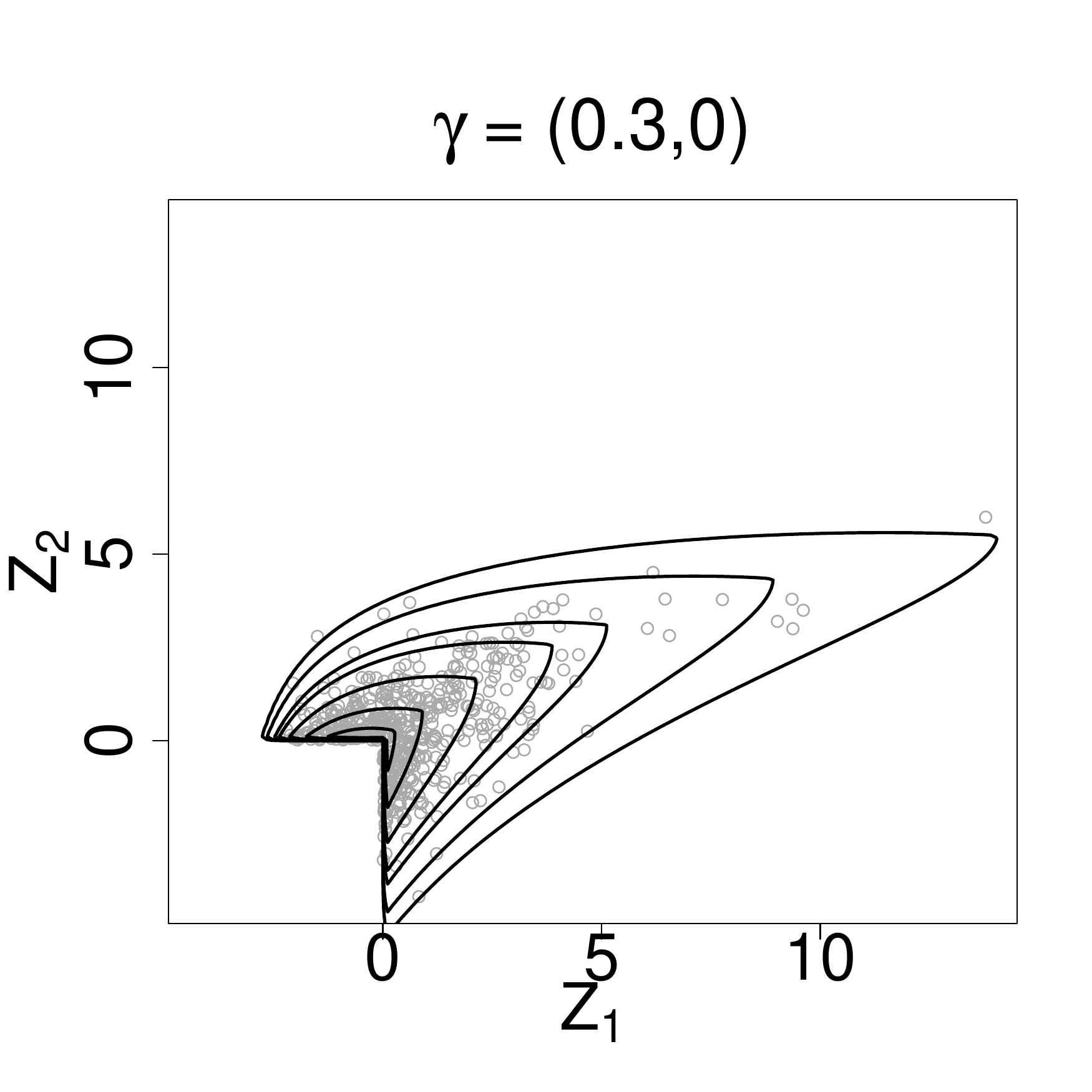}}
\subfloat{\includegraphics[scale=.35]{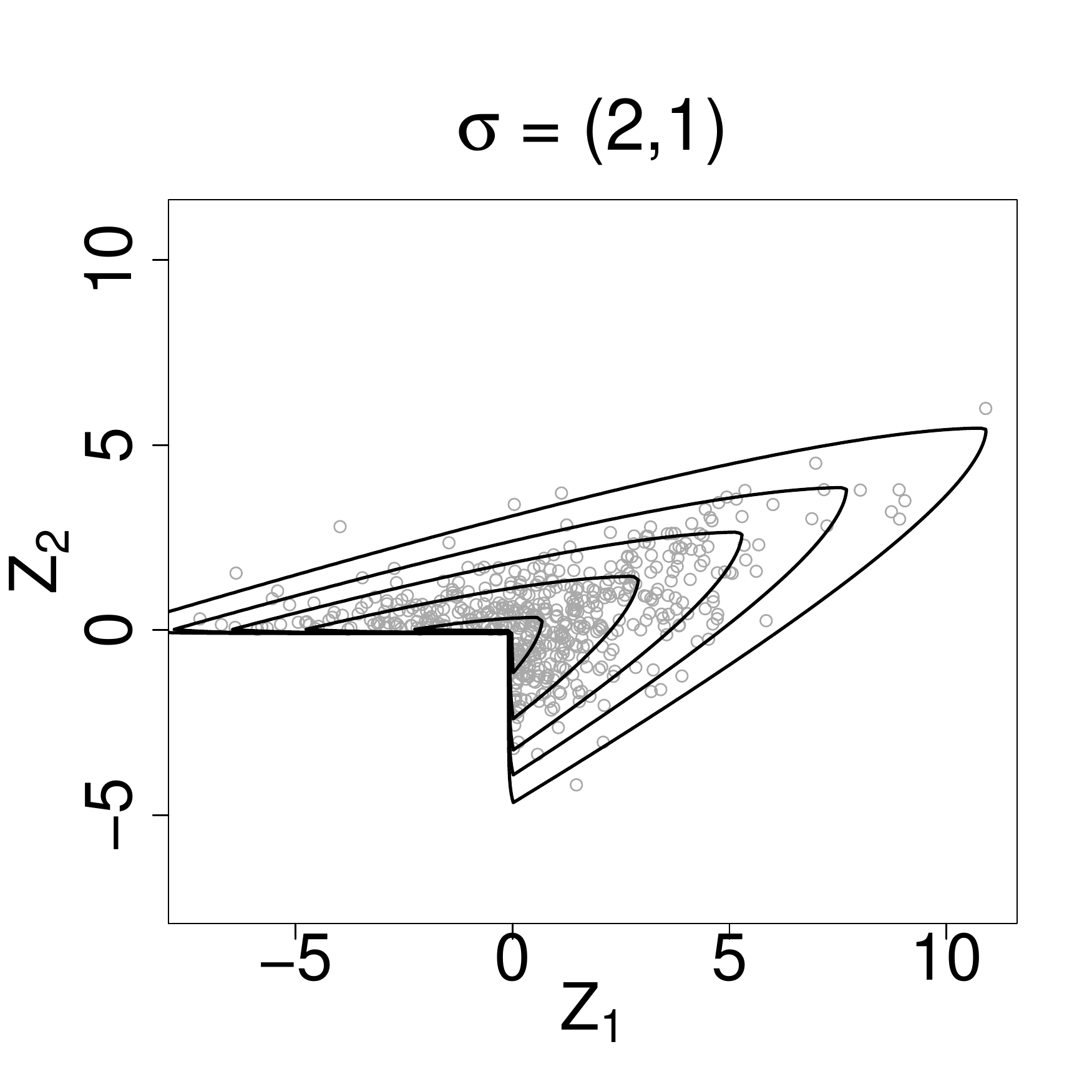}}
\caption{Scatterplots and density contours from 500 bivariate GPD random draws using \eqref{eq:Trep} and \eqref{eq:gpconstr} with parameters $\vc{\gamma} = (0.3,0)$, $\vc{\sigma} = (1,1)$ for the left panel and $\vc{\gamma} = (0,0)$, $\vc{\sigma} = (2,1)$ 
for the right panel. The generator $\vc{T}$ is zero-mean bivariate Gaussian with unit covariance matrix $I_2$. }
\label{fig:gauss}
\end{figure}

To make the link with the probabilities $p_0$ and $p_1$ used for EEA, we need a tool to project the information contained in  a possibly complex multivariate GPD structure into a single valued summary. The following proposition provides this key tool. 

\begin{prop}[Linear-projection, \citet{rootzen2018}] \label{prop:sumstab}
If $\vc{Z} \sim \textnormal{MGPD}(\vc{T},\vc{\sigma},\vc{\gamma})$ with $\vc{\gamma} = \gamma \vc{1}_d$, then for any non-negative weights $\vc{w} = (w_1,\ldots,w_d)^T$ such that $\PP[\vc{w}^T \vc{Z} > 0] > 0$,  the linear projection of $\vc{Z}$, conditioned on being positive, follows a univariate GPD, i.e.,
\begin{align*}
\left[  \vc{w}^T \vc{Z} \mid  \vc{w}^T \vc{Z}  > 0 \right]  \sim \textnormal{GPD} (\vc{w}^T \vc{\sigma}, \gamma). 
\end{align*}
\end{prop}

As explained in Section \ref{sec:introclimate}, our main inferential objective will be to estimate the probability $\PP[\vc{w}^T \vc{X} > v]$. A priori, if $\vc{X}$ represents a climatological index on $d$ grid points, it does not follow a multivariate GPD, but, since $\vc{X} - \vc{u} \mid \vc{X} \nleq \vc{u} \approx \vc{Z}$, we will use Proposition~\ref{prop:sumstab} to deduce a suitable model for $\vc{X}$ in Section \ref{sec:multivariate}.

\section{Causation probabilities for univariate extremes}\label{sec:univariate}
To understand first how PN, PS and  PNS behave for univariate extremes, we take
$
p_0(v) = \PP[X^{(0)} > v] 
$ 
and
$
p_1(v) = \PP[X^{(1)} > v] 
$ 
when $X^{(0)}$ and $X^{(1)}$ are either Gaussian or GPD random variables. The left panel of 
Figure \ref{fig: far Gauss-Pareto} 
shows the case where the counterfactual world corresponds to a standardized Gaussian variable, $X^{(0)} \sim N(0,1)$, and the factual world 
is one Kelvin warmer with a higher variability, $X^{(1)} \sim N(1,1.5)$. This artificial design mimics the typical behaviour of  mean temperature anomalies. 
Two features can be highlighted from this example:
PN goes to one as $v$ increases, and the maximum of PNS is around $0.35$. In other words, the probability of necessary causation becomes certain for extremes (large $v$), and the probability of necessary and sufficient causation can be reasonably high in the Gaussian case.
\begin{figure}[ht!]
\begin{center}
\includegraphics[scale=.255]{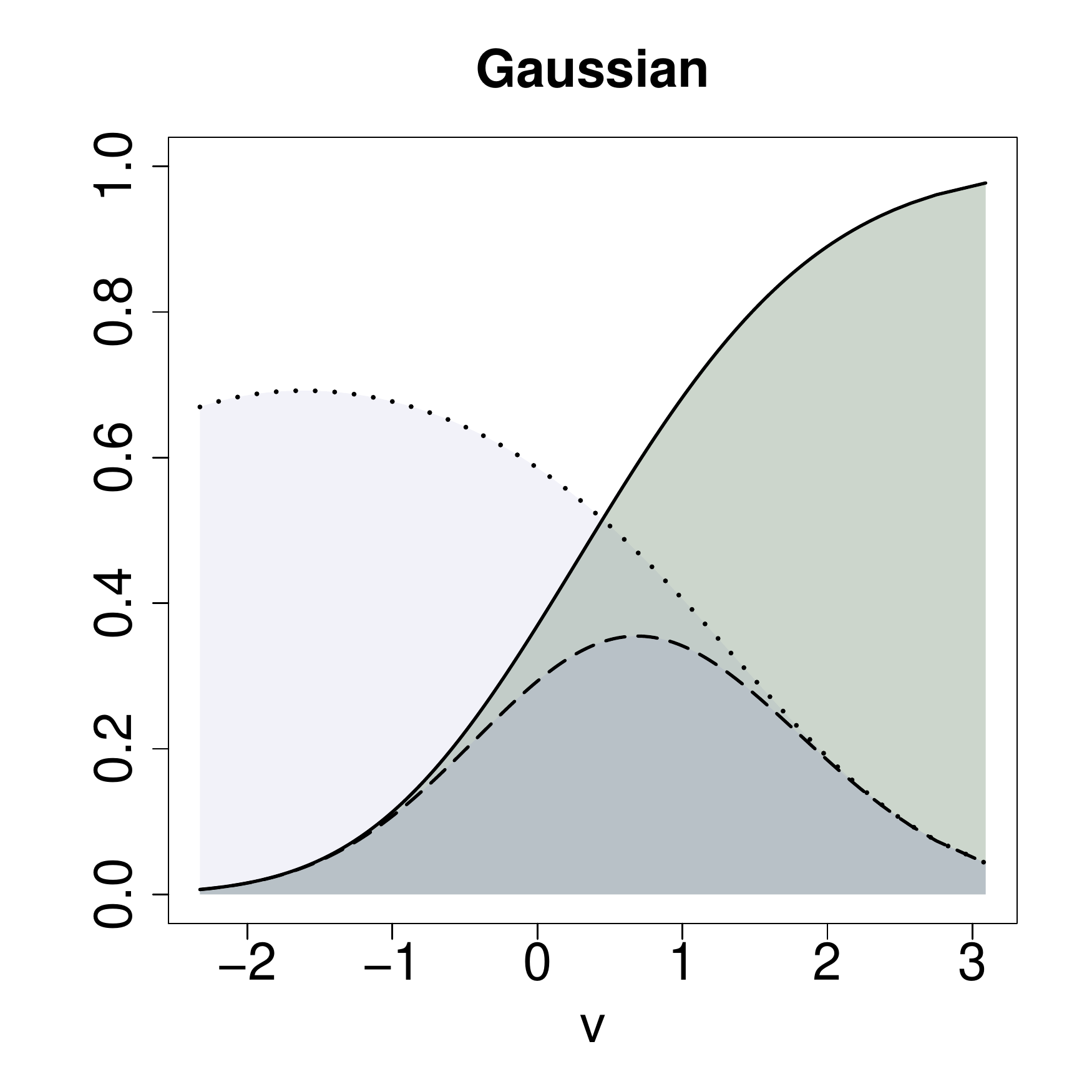}  \hspace{-0.7cm}
\includegraphics[scale=.255]{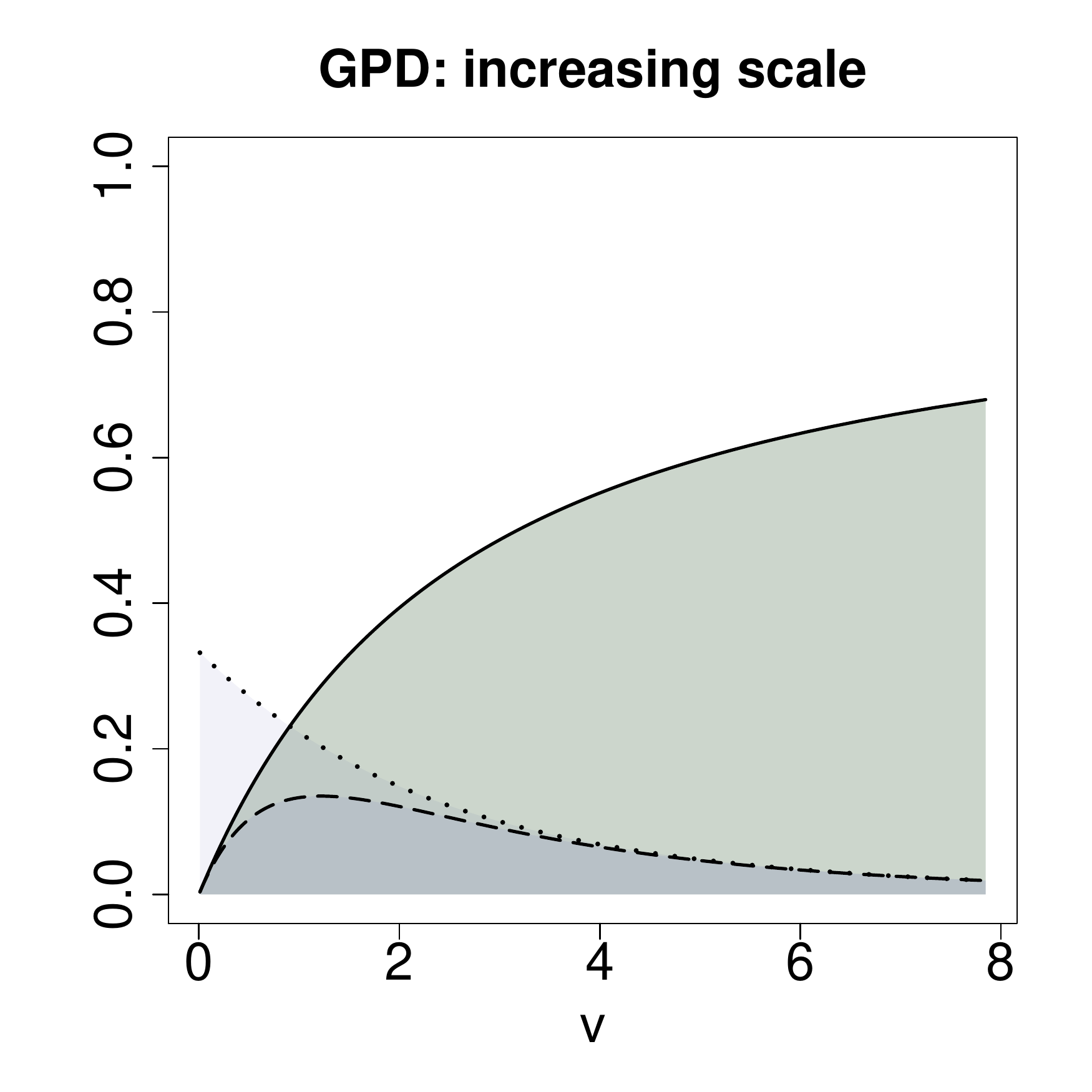}
 \hspace{-0.7cm}
\includegraphics[scale=.255]{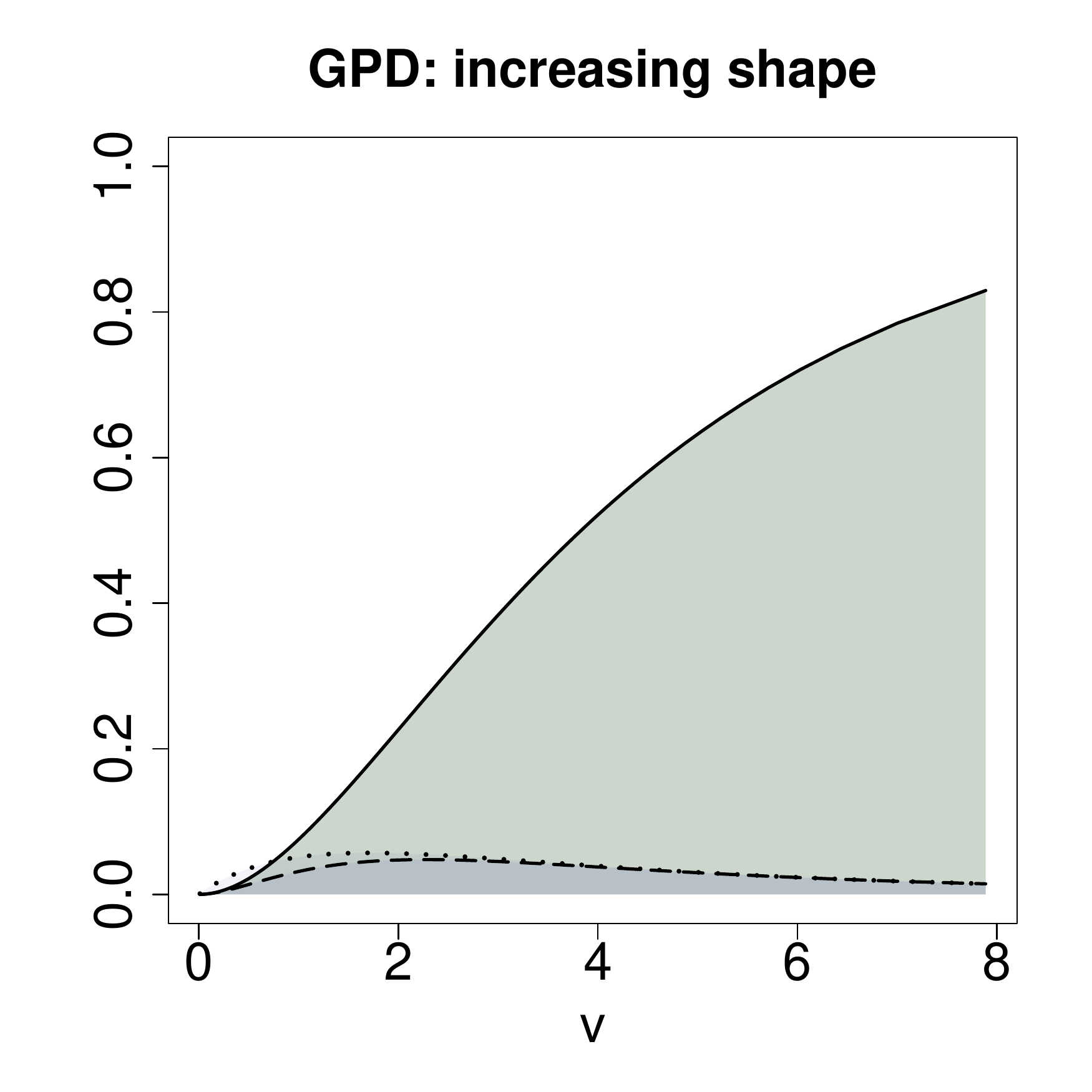}
\caption{Probabilities of necessary causation (PN, solid line), sufficient causation (PS, dotted line) and necessary and sufficient causation (PNS, dashed line) as functions of $v$. 
Left panel: Gaussian set-up, $N(0,1)$ for the counterfactual world and $N(1,1.5)$ for the factual one.  
Middle and right panels: GPD set-up, GPD$(1,0.2)$ for the counterfactual world, GPD$(1.5,0.2)$ for the factual one (middle) and GPD$(1,0.1)$ for the counterfactual world, GPD$(1,0.3)$ for the factual one (right).}
\label{fig: far Gauss-Pareto}
\end{center}
\end{figure}

To contrast these remarks with other types of tail behaviors, 
the middle panel of Figure \ref{fig: far Gauss-Pareto} displays a GPD case with equal shape parameter $\gamma=0.2$ in the counterfactual and factual worlds, but different scale parameters, $\sigma^{(0)}=1$ and $\sigma^{(1)}=1.5$.  
One can see that, as $v$ increases,  PN converges to a constant around $0.7<1$, and PNS remains small for any value of $v$. Hence, causal evidencing is much more difficult than in the Gaussian set-up, where a rare event in the factual world ($p_1$ small) would be nearly impossible in the counterfactual world ($p_0$ almost zero). 
In contrast, even a very rare event in the factual world will not be impossible in a GPD counterfactual world. 
Concerning PNS, it is small in the second panel and this phenomenon is even more pronounced when the shape parameter changes between the two worlds; see the right panel where $\gamma^{(0)}=0.1$, $\gamma^{(1)}=0.3$, and $\sigma^{(0)}=\sigma^{(0)}=1$.
As PNS is near zero for all $v$, there is no reason to maximize it. Instead, maximizing causality will correspond to maximizing  PN in the remainder of this work. 

In practice, $X^{(0)}$ and $X^{(1)}$ do not follow GPDs. Using a classical peaks-over-thresholds approach, we can condition on some high threshold $u^{(i)}$ and approximate the probabilities $p_i (v) = \PP[X^{(i)} > v]$ for $v > u^{(i)} $ by
\begin{equation}\label{eq:puni}
p_i (v) \approx \PP[X^{(i)} > u^{(i)}] \, \overline{H} \left( v - u^{(i)} ; \sigma^{(i)}, \gamma^{(i)} \right), \qquad  \text{ for } i \in \{0, 1\}.
\end{equation}
We can now formalize the tail behaviour observed in Figure \ref{fig: far Gauss-Pareto}. Whenever the limit of PN($v$) for large $v$ is finite\footnote{Degenerate cases can occur when $p_1(v)=0$.
For example, if $\gamma^{(1)} < 0$, the PN is not defined for $v \geq u^{(1)} - \sigma^{(1)}/\gamma^{(1)}$, which is visible for the dashed line in the left panel of Figure \ref{fig: FARuni}.}, it has to  be equal to 
\begin{equation}\label{eq:pnlimit}
\begin{dcases}
\mathbbm{1} \left\{ \gamma^{(0)} < \gamma^{(1)} \right\} & \text{ if } \gamma^{(1)} \neq \gamma^{(0)}, \\
1 - \frac{p_0 (u^{(0)})}{p_1 (u^{(1)})} \left(\frac{\sigma^{(0)}}{\sigma^{(1)}}\right)^{1/\gamma} & \text{ if } \gamma^{(1)} = \gamma^{(0)} =: \gamma, \, \gamma \neq 0, \\
\mathbbm{1} \left\{ \sigma^{(0)} < \sigma^{(1)} \right\} & \text{ if } \gamma^{(1)} = \gamma^{(0)} = 0.
\end{dcases}
\end{equation} 
where $\mathbbm{1}(A)$ represents the indicator function, equal to one if $A$ is true and zero otherwise.

The left panel of Figure \ref{fig: FARuni} shows how three different GPD shape parameters, $\gamma = -0.4$, $\gamma = 0$ and $\gamma = 0.4$ (dashed, solid and dotted lines respectively) influence the increase of the PN with respect to the threshold $v$. The right panel of Figure \ref{fig: FARuni} points out possible atypical behaviours of the PN, highlighting
that PN$(v)$ is not always increasing as $v$ increases. Here, 
the solid line corresponds to a counterfactual world with $(\gamma^{(0)},\sigma^{(0)}) = (0,2)$  compared  to  a factual world with $(\gamma^{(1)}, \sigma^{(1)} )= (0.4, 1)$, while the dotted line represents the converse change: from $(\gamma^{(0)},\sigma^{(0)}) = (0.4,1)$  to $(\gamma^{(1)}, \sigma^{(1)} )= (0,2$).

\begin{figure}[ht!]
\centering
\includegraphics[scale=.36]{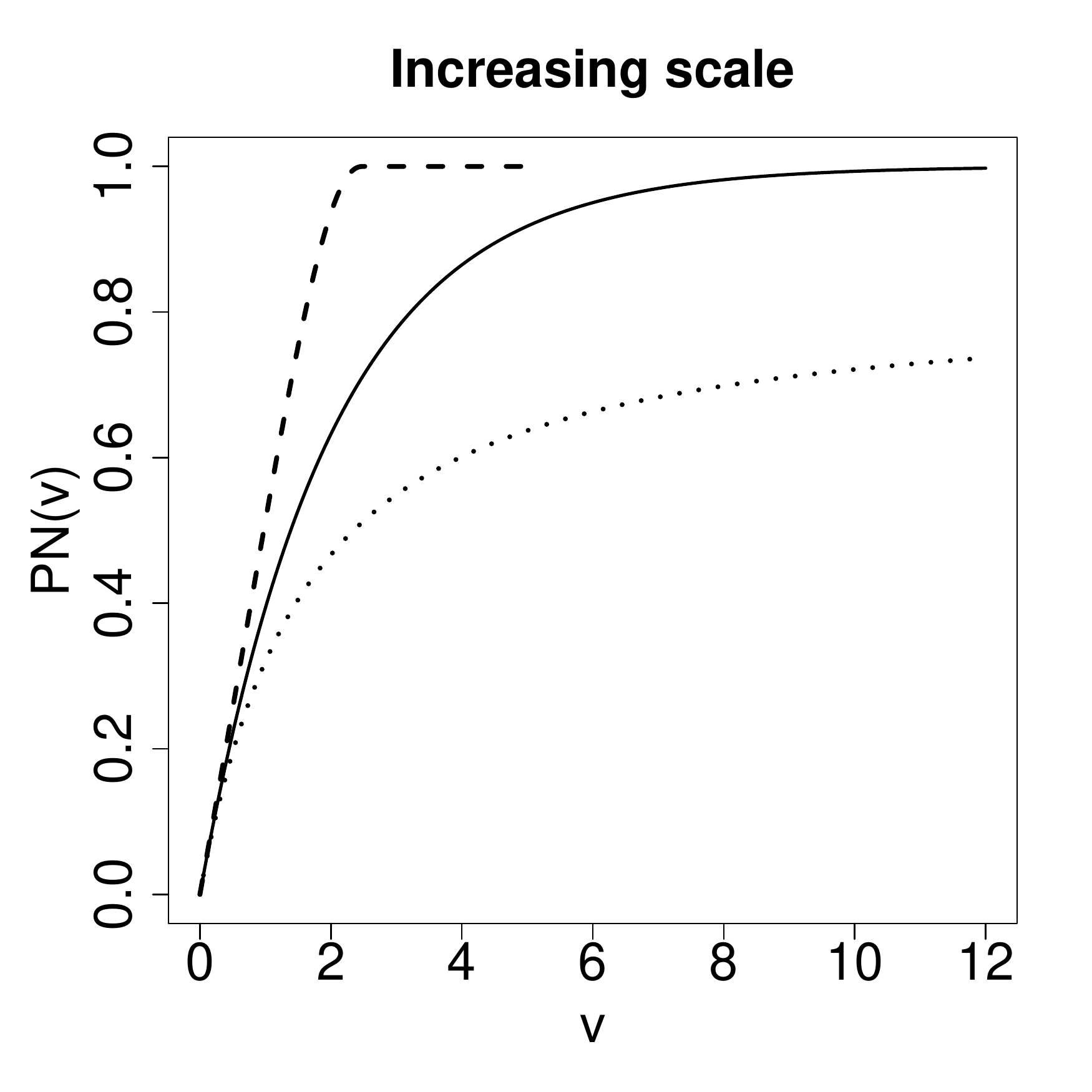} 
\hspace{-0.5cm}
\includegraphics[scale=.36]{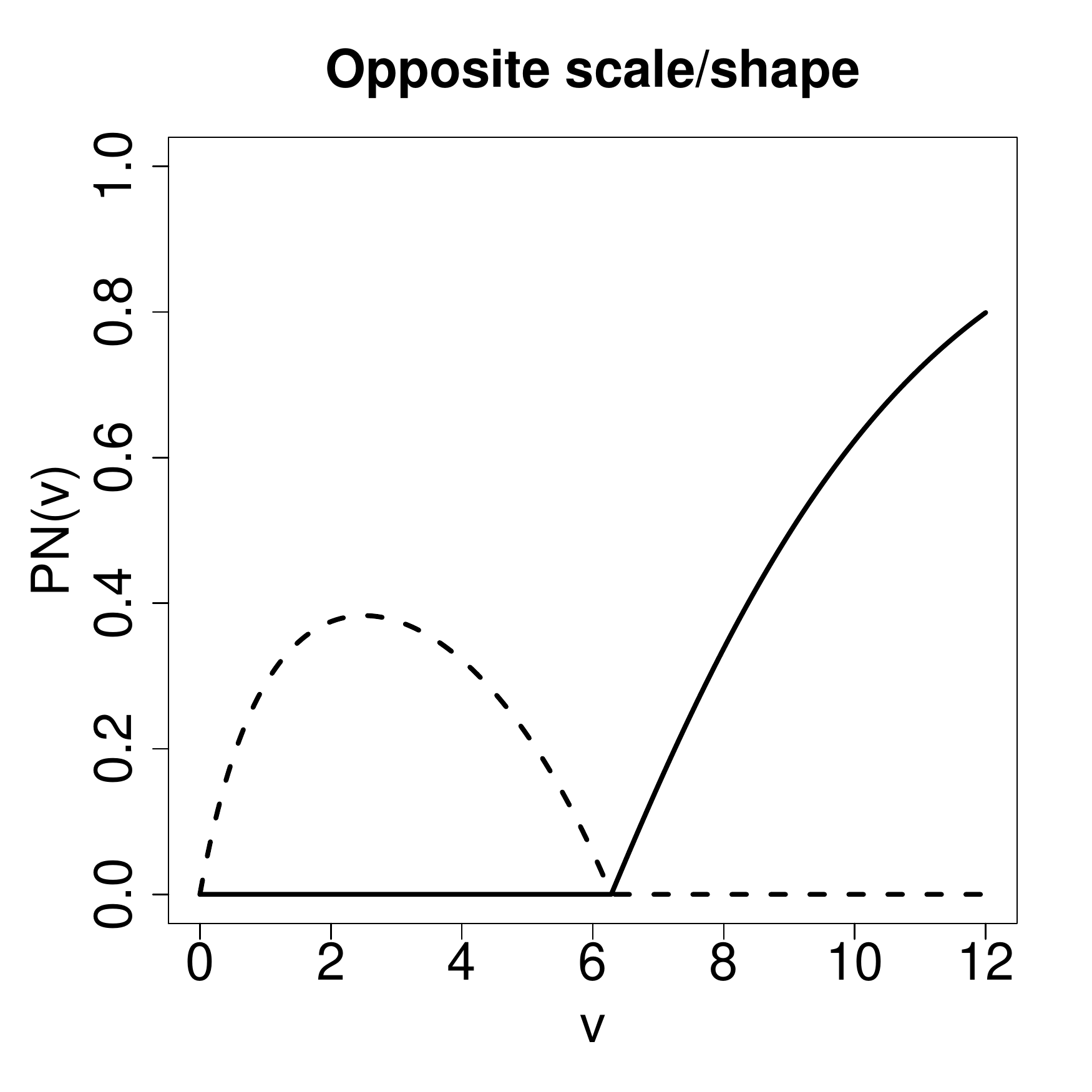}
\caption{Probability of necessary causation as a function of $v$, see (\ref{eq:probas}) and (\ref{eq:puni}). 
Left panel: the counterfactual scale is $\sigma^{(0)} = 1$, increasing to $\sigma^{(1)} = 2$ in the factual world, while the shape parameter is identical, $\gamma^{(0)}=\gamma^{(1)}$, equal to  $-0.4$, $0$, and $0.4$ for the dashed, solid and dotted lines respectively.
Right panel: the solid line  corresponds to increasing shape and decreasing scale, $(\gamma^{(0)},\sigma^{(0)}) = (0,2)$  and $(\gamma^{(1)}, \sigma^{(1)} )= (0.4, 1)$. 
The dashed line corresponds to the opposite scenario : $(\gamma^{(0)},\sigma^{(0)}) = (0.4,1)$  and $(\gamma^{(1)}, \sigma^{(1)} )= (0,2$).}
\label{fig: FARuni}
\end{figure}
%

\section{Necessary causation in a multivariate set-up}\label{sec:multivariate}

\subsection{The multivariate GPD and necessary causation}
Let $\vc{X}$ be any $d$-dimensional random vector such that $\left[ \vc{X} - \vc{u} \mid \vc{X} \nleq \vc{u}  \right]  \approx \vc{Z}$ for some high threshold $\vc{u} \in \RR^d$, where 
$\vc{Z} \sim \textnormal{MGPD}(\vc{T},\vc{\sigma},\vc{\gamma})$. 
Then, according to Proposition \ref{prop:sumstab}, the extremal information contained in any linear projection $\vc{w}^T \vc{X}$ can be approximated, up to a normalizing constant, by a univariate GPD survival function. More precisely, for any $v > \vc{w}^T \vc{u}$, we can write   
\begin{align}\label{eq:pmulti2}
\PP[\vc{w}^T \vc{X} > v] & = \PP[\vc{w}^T \vc{X} > \vc{w}^T \vc{u} ]  \, \PP[\vc{w}^T (\vc{X} - \vc{u}) > v - \vc{w}^T \vc{u} \mid \vc{w}^T (\vc{X}  - \vc{u}) > 0] \notag\\
& \approx \PP \left[ \vc{w}^T \vc{X} > \vc{w}^T \vc{u}  \right] \, \overline{H} \left( v - \vc{w}^T \vc{u} ; \vc{w}^T \vc{\sigma}, \gamma \right), 
\end{align}
for $w_1+ \dots+ w_d  = 1$ and $\vc{\gamma} = \gamma \vc{1}_d$. Constraining the weights to sum to a constant is necessary to ensure identification of $\vc{\sigma}$. The condition $\vc{\gamma} = \gamma \vc{1}_d$ implies that conditional marginal distributions $\left[ Z_j \mid Z_j >  0 \right] $ have equal shape parameters for $j=1,\dots,d$. 
Therefore, homogeneous spatial regions (in terms of the shape parameter) have to be identified in practice. 
This is closely related to the regional frequency analysis problem treated in hydrology \cite[][]{carreau2017}. Finally, we note that the dependence structure of $\vc{X}$ is present in the term $\PP \left[ \vc{w}^T \vc{X} > \vc{w}^T \vc{u}  \right]$ only.

Any linear projection in the factual and counterfactuals worlds, denoted $p_1 (v ; \vc{w}) = \PP [\vc{w}^T \vc{X}^{(1)} > v]$ and $p_0 (v ; \vc{w}) = \PP [\vc{w}^T \vc{X}^{(0)} > v]$ respectively, can now be used to compute a probability of necessary causation that depends on the weight $\vc{w}$ and the dependence structure of $\vc{X}^{(0)}$ and $\vc{X}^{(1)}$,
\begin{equation}\label{eq:pn-multi}
\textnormal{PN}(v,  \vc{w})  = \max \left( 1 - \frac{\PP[\vc{w}^T \vc{X}^{(0)} > v]  }{\PP[\vc{w}^T \vc{X}^{(1)} > v] } , 0\right), \,\,\,
\end{equation}
where 
$\vc{X}^{(i)}$ satisfies approximation (\ref{eq:pmulti2}) for $i \in \{0,1 \}$. 
To understand how dependence affects the strength of necessary causation, we study the value of $\textnormal{PN}(v,  \vc{w})$ in the bivariate  case with $\vc{w}=(0.5,0.5)$,  
$ \vc{X}^{(0)} \overset{\mathrm{d}}{=} \vc{Z}^{(0)} \sim \textnormal{MGPD}\left(\vc{T}^{(0)},\vc{\sigma}^{(0)},\vc{0}\right)$ 
and $ \vc{X}^{(1)}  \overset{\mathrm{d}}{=} \vc{Z}^{(1)} \sim \textnormal{MGPD}(\vc{T}^{(1)},\vc{\sigma}^{(1)},\vc{0})$. 
In the example displayed in Figure \ref{fig:FARmulti}, 
two different dependence structures are investigated, summarized by the tail dependence coefficient $\chi$.
The dotted line corresponds to increasing dependence, from $\chi^{(0)}= 0.3$ to $\chi^{(1)}=0.5$ and an increasing marginal scale, from $\vc{\sigma}^{(0)} = (1,1)$ to $\vc{\sigma}^{(1)} = (2,2)$. The dashed line again represents increasing dependence, but decreasing marginal scale, from $\vc{\sigma}^{(0)} = (2,2)$ to $\vc{\sigma}^{(1)} = (1,1)$. Finally, the solid line shows increasing marginal scale of the same order as the dotted line, and decreasing dependence, from $\chi^{(0)}= 0.5$ to $\chi^{(1)}=0.3$.
\begin{figure}[ht!]
\centering
\subfloat{\includegraphics[scale=.36]{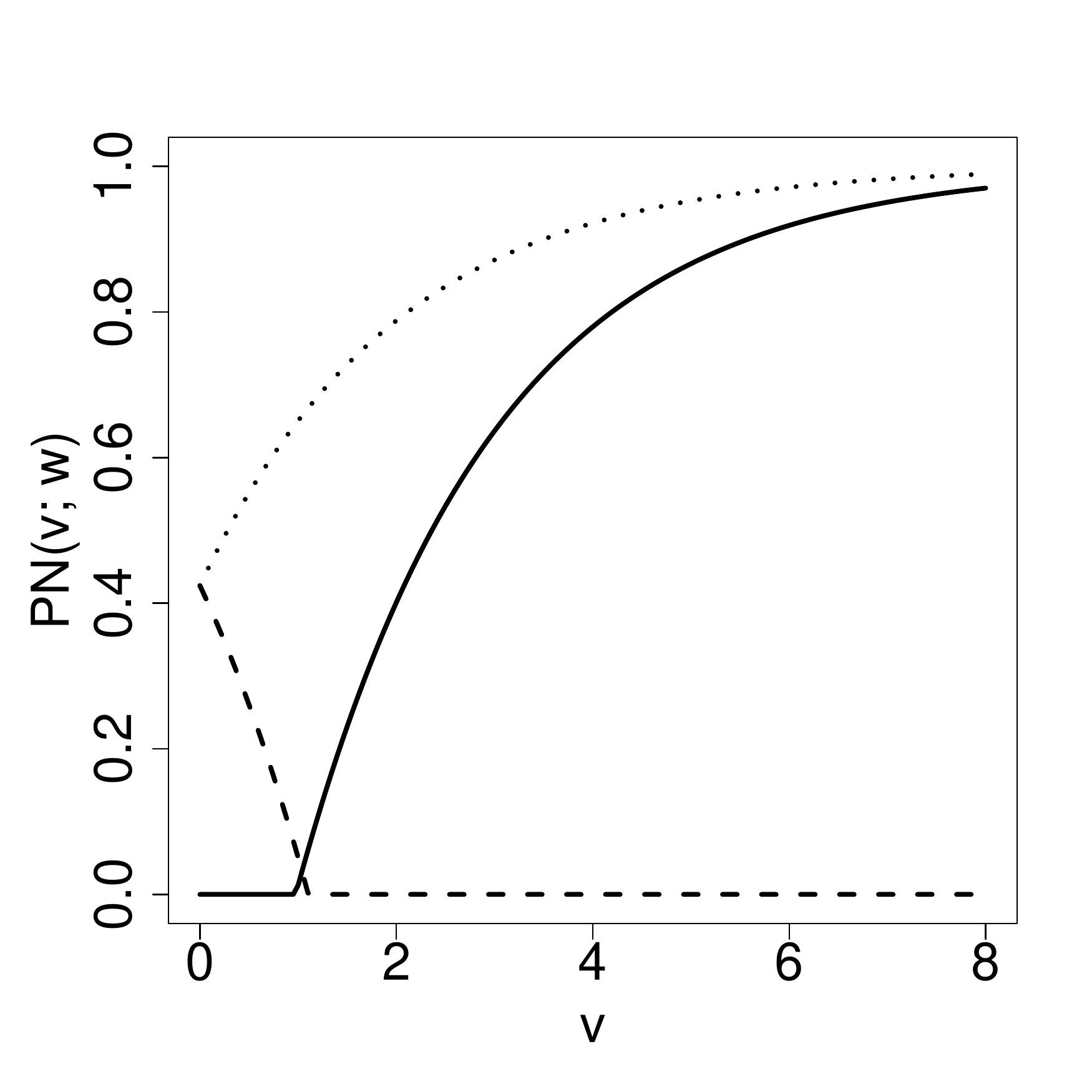}}
\caption{
PN$(v,\vc{w}=(0.5,0.5))$ defined by (\ref{eq:pn-multi}) between two bivariate GPDs,
$\vc{Z}^{(0)} \sim \textnormal{MGPD}(\vc{T}^{(0)},\vc{\sigma}^{(0)},\vc{0})$ 
and $\vc{Z}^{(1)} \sim \textnormal{MGPD}(\vc{T}^{(1)},\vc{\sigma}^{(1)},\vc{0})$.
The dotted line corresponds to $\chi^{(0)}= 0.3$, $\chi^{(1)}=0.5$, $\vc{\sigma}^{(0)} = (1,1)$ and $\vc{\sigma}^{(1)} = (2,2)$. The dashed line differs from the dotted line by $\vc{\sigma}^{(0)} = (2,2)$ and $\vc{\sigma}^{(1)} = (1,1)$. The solid line differs from the dotted line by $\chi^{(0)}=0.5$ and $\chi^{(1)}=0.3$.}
%
%
\label{fig:FARmulti}
\end{figure} 
Figure \ref{fig:FARmulti} shows that the dependence structure can have an impact on the PN for any finite value of $v$. In other words, EEA based on a hypothesis of independence (eg, in space) will lead  
to incomplete statements concerning the strength of PN whenever the multivariate extremes are dependent. Figure \ref{fig:FARmulti} also suggests that, as $v$ increases, the impact of an increasing dependence in the factual world becomes negligible. However, it is important to keep in mind that in applications, the marginal scale might be constant between the two worlds. In that case, we will see the impact of increasing dependence in Figure \ref{fig:univsmulti}.

\subsection{Maximizing necessary causation}\label{sec:weights}
In a multivariate Gaussian set-up, \cite{hannart18} proposed to maximize causation probabilities by using the linear projection that will contrast the factual and counterfactual worlds the most. Their solution was similar to linear discriminant analysis. This leads to the question of how to reduce the dimension of a multivariate GPD vector, while ensuring that the projected data contains the most information in terms of causality for extremes.

More specifically, the choice of  $\vc{w}$ plays an essential role in the maximization of necessary causation for multivariate GPD random variables. To address this point in the bivariate case, we need the following result. 
 \begin{prop}\label{prop:optpn}
 Let $\gamma \in \RR$ and consider two positive bivariate scale parameters: $\vc{\sigma}^{(0)} =  (\sigma_1^{(0)},\sigma_2^{(0)})^T$ and $\vc{\sigma}^{(1)} =(\sigma_1^{(1)},\sigma_2^{(1)})^T$.
 Denote  
 $$
 R = \frac{\left( \sigma_1^{(0)}\sigma_2^{(1)} - \sigma_2^{(0)}\sigma_1^{(1)}\right) \left(  \sigma_1^{(0)}\sigma_2^{(1)} - \sigma_2^{(0)}\sigma_1^{(1)}  + \gamma v \left\{ (\sigma_2^{(1)} - \sigma_1^{(1)}) - ( \sigma_2^{(0)} - \sigma_1^{(0)}) \right\} \right)}{   (\sigma_2^{(1)} - \sigma_1^{(1)}) ( \sigma_2^{(0)} - \sigma_1^{(0)})   \left\{ (\sigma_2^{(1)} - \sigma_1^{(1)}) - ( \sigma_2^{(0)} - \sigma_1^{(0)}) \right\}^2 },
 $$
and if $R>0$, define the weights  
 $$
 w_{\pm} (v)   =  \frac{ \sigma_2^{(1)} - \sigma_2^{(0)}}{(\sigma_2^{(1)} - \sigma_1^{(1)}) - (\sigma_2^{(0)} - \sigma_1^{(0)})}   \pm 
\sqrt{R}. 
$$
 If $R  \geq 0$ and if one of the two weights $w_{\pm} (v) $ belongs to $(0,1)$, then this weight, denoted $w_{\textnormal{opt}}$, maximizes 
\begin{equation}\label{eq:ratio}
 \left\{ 1 - \frac{ \overline{H} \left( v ; \,w \sigma_1^{(0)} +(1-w) \sigma_2^{(0)}, \gamma \right)}{\overline{H} \left( v ; w \sigma_1^{(1)} +(1-w) \sigma_2^{(1)}, \gamma \right)} \right\}. 
\end{equation}
In all other cases, only zero or unit weights maximize this ratio. 
\end{prop}
When $\gamma = 0$, $w_{\pm} (v)$ is simpler because it does not depend on $v$. 
Expression \eqref{eq:ratio} is an approximation of the PN defined in \eqref{eq:pn-multi}; it is equal to the PN when $\vc{X}^{(0)}, \vc{X}^{(0)}$ are multivariate GPDs and when $\PP [\vc{w}^T \vc{Z}^{(0)} > 0] = \PP [\vc{w}^T \vc{Z}^{(1)} >0]$, i.e., when the dependence structure remains constant between the two worlds.
Proposition \ref{prop:optpn} allows us to study the gain in terms of PN with respect to the weight $w$. 
When unit or zero weights are chosen as the optimal solution in Proposition \ref{prop:optpn}, 
only one coordinate is considered and no linear projection is necessary. This happens when the contrast in one of the margins between the factual and counterfactual world is already sufficient to optimize PN. 
However, Proposition \ref{prop:optpn} shows that, to maximize necessary causality, one needs to consider only those components that (individually) give the largest PN. As an example, take $\vc{X}^{(0)} \overset{\mathrm{d}}{=} \vc{Z}^{(0)} \sim \textnormal{MGPD}(\vc{T}^{(0)}, (1,2)^T, \gamma \vc{1}_d)$ 
and $\vc{X}^{(1)} \overset{\mathrm{d}}{=} \vc{Z}^{(1)} \sim \textnormal{MGPD}(\vc{T}^{(1)},(1.5,2)^T,\gamma \vc{1}_d)$. The dependence structures of $\vc{T}^{(1)}$ and $\vc{T}^{(0)}$ are chosen such that $\chi^{(0)} = \chi^{(1)} = 0.5$. 
Hence, the difference between the two worlds is only due to the scale change in one of the components. 
Figure \ref{fig:multi} shows the PN gain as a function of $v$, i.e., the ratio between 
$\textnormal{PN}(v,  (0.5,0.5)^T)$ and $\textnormal{PN}(v,  (w_{\textnormal{opt}},1-w_{\textnormal{opt}})^T)$ where $w_{\textnormal{opt}} = 1$ based on Proposition \ref{prop:optpn}. 
Each curve corresponds to a different shape parameter (with constraint $\gamma^{(0)} = \gamma^{(1)}$), equal to  $-0.4$ (dashed line), $0$ (solid line), and $0.4$ (dotted line), respectively. We see that the optimal weight can lead to a large increase in necessary causality, particularly when the shape parameter is positive (dotted line).

\begin{figure}[ht!]
\centering
\subfloat{\includegraphics[scale=.36]{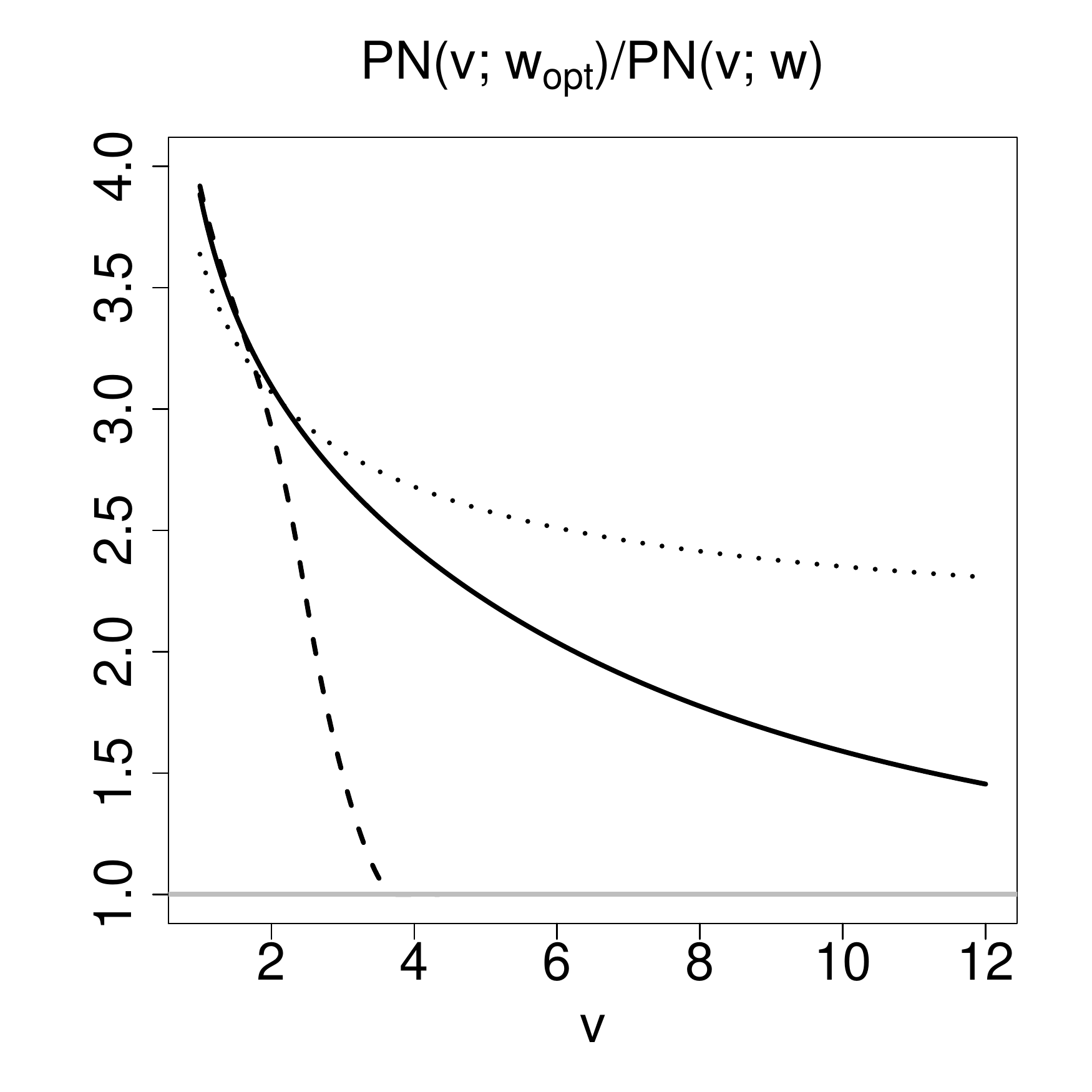}}
\caption{
Necessary causation gain for $\vc{X}^{(0)} \overset{\mathrm{d}}{=} \vc{Z}^{(0)} \sim \textnormal{MGPD}(\vc{T}^{(0)}, (1,2)^T, \gamma \vc{1}_d)$ 
and $\vc{X}^{(1)} \overset{\mathrm{d}}{=} \vc{Z}^{(1)} \sim \textnormal{MGPD}(\vc{T}^{(1)},(1.5,2)^T,\gamma \vc{1}_d)$, where $\vc{T}^{(1)}$, $\vc{T}^{(0)}$ are Gaussian random vectors such that $\chi^{(0)} = \chi^{(1)} = 0.5$. The ratio of $\textnormal{PN}(v,  (w_{\textnormal{opt}},1-w_{\textnormal{opt}})^T)$ to $\textnormal{PN}(v,  (0.5,0.5)^T)$ is shown as a function of $v$, where $w_{\textnormal{opt}} = 1$ based on Proposition \ref{prop:optpn}. The dashed, solid and dotted lines correspond to a shape parameter (with constraint $\gamma^{(0)} = \gamma^{(1)}$) of  $-0.4$, $0$ and $0.4$ respectively.
}
\label{fig:multi}
\end{figure} 

Explicit optimal weights like in Proposition \ref{prop:optpn} can only be obtained in very specific cases. For example, for the bivariate Gaussian GPD with $\gamma = 1$, the probability $\PP[\vc{w}^T \vc{Z}^{(i)} > 0]$ does not depend on $\vc{w}$ (see Proposition A.1 in the supplementary material, \citet{KirilioukNaveau20}). For most other cases, numerical optimization schemes have to be used, especially beyond the bivariate set-up. In order to move closer to practical applications, we need to couple this optimization procedure with inference in a multivariate context.

 
\subsection{Inference}

Let $\vc{X}^{(0)}_1,\ldots,\vc{X}^{(0)}_n$ and $\vc{X}^{(1)}_1,\ldots,\vc{X}^{(1)}_n$ denote two independent samples of size $n$, representing climate model output in the counterfactual and the factual world respectively, and let $\vc{u}^{(0)}$, $\vc{u}^{(1)}$ denote two high thresholds.  For $i \in \{0,1 \}$, let $N_i$ denote the number of observations among $\vc{X}^{(i)}_1,\ldots,\vc{X}^{(i)}_n$ that have at least one component exceeding $\vc{u}^{(i)}$. Extracting these observations and subtracting $\vc{u}^{(i)}$, we obtain the multivariate GPD samples $\vc{Z}_1^{(i)},\ldots,\vc{Z}^{(i)}_{N_i}$.
For $v > \vc{w}^T \vc{u}^{(i)}$, an estimator of $p_i (v ; \vc{w}) = \PP [ \vc{w}^T \vc{X}^{(i)} > v]$ and hence of the PN follows from approximation \eqref{eq:pmulti2}. The first term, $ \PP \left[ \vc{w}^T \vc{X}^{(i)} > \vc{w}^T \vc{u}^{(i)}  \right]$, can be estimated nonparametrically by 
\begin{equation*}
\widehat{p}_i^{(\emp)} (v; \vc{w}) = \frac{1}{n} \sum_{t=1}^n \mathbbm{1} \left\{ \vc{w}^T \vc{X}_t^{(i)} > v \right\}.
\end{equation*}
To estimate the second term, $\overline{H} \left(v - \vc{w}^T \vc{u}^{(i)} ; \vc{w}^T \vc{\sigma}^{(i)},\gamma^{(i)} \right)$, we first compute estimators $\widehat{\vc{\sigma}}^{(i)}$ and $\widehat{\vc{\gamma}}^{(i)}$ by applying the method of probability weighted moments to $(Z^{(i)}_{tj} \mid Z^{(i)}_{tj} > 0)_{t = 1,\ldots,N_i}$ for $j \in \{1,\ldots,d\}$ (see Section C in the supplementary material, \citet{KirilioukNaveau20}). Next, we set $\widehat{\gamma}^{(i)} = d^{-1} \sum_{j=1}^d \widehat{\gamma}_j^{(i)}$ \footnote{An alternative method to enforce equal shape parameters is described in \citet{carreau2017}}. 
Finally, we estimate $p_i (v; \vc{w})$ by
\begin{multline}\label{eq:pest2}
\widehat{p}_i (v ;\vc{w}) = \\
\begin{dcases}
\widehat{p}_i^{(\emp)} (v ; \vc{w}) & \text{ if } v \leq \vc{w}^T \vc{u}^{(i)}, \\
\widehat{p}_i^{(\emp)} (\vc{w}^T \vc{u}^{(i)} ; \vc{w}) \, \overline{H} \left(v - \vc{w}^T \vc{u}^{(i)} ; \vc{w}^T \vc{\widehat{\sigma}}^{(i)},\widehat{\gamma}^{(i)} \right) & \text{ if } v \leq \vc{w}^T \vc{u}^{(i)}. \\
\end{dcases}
\end{multline}
Alternatively, we could directly estimate $\gamma^{(i)}$ and $\vc{w}^T \vc{\sigma}^{(i)}$ by applying the method of probability weighted moments to $(\vc{w}^T \vc{Z}_t^{(i)} \mid \vc{w}^T \vc{Z}^{(i)}_t > 0)_{t = 1,\ldots,N_i}$, which reduces uncertainty and enforces the constraint of equal shape parameters, but requires the weights $\vc{w}$ to be chosen upfront.
Section D in the supplementary material \citep{KirilioukNaveau20} shows a small simulation experiment, confirming the good performance of $\widehat{\textnormal{PN}} = 1 -\widehat{p}_0 / \widehat{p}_1$.

In the previous sections, we studied the increase in PN for changing dependence structures and marginal parameters. Another important question is what happens when marginal parameters do not change ($\vc{\sigma}^{(0)} = \vc{\sigma}^{(1)}$ and $\vc{\gamma}^{(0)} = \vc{\gamma}^{(1)}$), while dependence increases ($\chi^{(1)} > \chi^{(0)}$). 
Under a hypothesis of independence in space, one would aggregate the observations from all grid points  (i.e., calculate $\vc{w}^T \vc{X}^{(i)}$) and estimate the univariate PN, thus possibly underestimating the true PN.
To see by how much, and how the result varies with the dimension, we conduct the following experiment. Consider $d = 9$ points on a regular $3 \times 3$ unit distance grid. For distances from $1$ to $\sqrt{8}$, pairwise tail dependence coefficients ranging from $0.4$ to $0.3$ for the counterfactual world and from $0.55$ to $0.4$ for the factual world were obtained using a Whittle-Mat\'ern correlation function\footnote{The covariance matrices $\Sigma^{(0)}$ and $\Sigma^{(1)}$ are generated using a Whittle-Mat\'ern correlation function with fixed shape $\kappa^{(0)} = \kappa^{(1)} = 1$ and varying scales $\phi^{(0)} = 1$, $\phi^{(1)} = 2.5$. The correlation matrices are then multiplied by 10 to obtain $\Sigma^{(0)}$ and $\Sigma^{(1)}$}.  
We evaluate the PN in the $99 \%$ quantile of $\vc{w}^T\vc{Z}^{(0)}$ using equal weights, calculated based on a pre-simulation run of sample size $10^6$ and held fixed. Figure~\ref{fig:univsmulti} shows boxplots of the multivariate estimates $\widehat{\textnormal{PN}}$ minus the univariate estimates, based on 1000 samples of size $n = 2000$. The black line corresponds to the true values, calculated using the formulas in Sections A and B in the supplementary material \citep{KirilioukNaveau20}. We see that as the dimension increases, taking dependence into account increases necessary causation.
\begin{figure}[ht!]
\centering
\subfloat{\includegraphics[scale=.36]{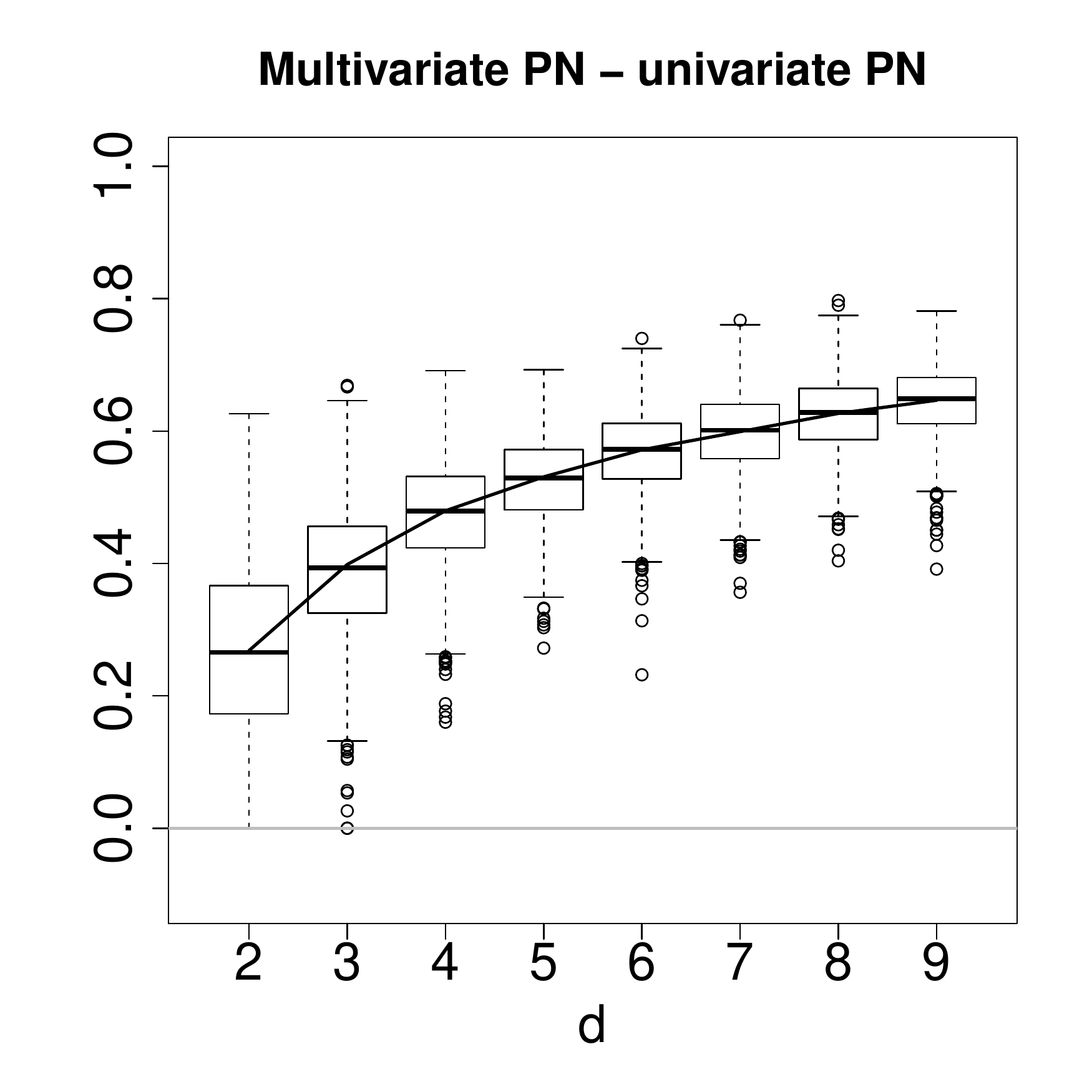}}
\caption{Boxplots of the multivariate estimates $\widehat{\textnormal{PN}}  = 1 -\widehat{p}_0 / \widehat{p}_1$ minus the univariate PN estimates of aggregated data, where $\widehat{p}_i$ is defined in \eqref{eq:pest2}, and $d \in \{2,\ldots,9\}$. 1000 samples of size $n = 2000$ were simulated from a multivariate Gaussian GPD model with $\vc{\sigma}^{(0)} = \vc{\sigma}^{(1)} = 1$ and $\vc{\gamma}^{(0)} = \vc{\gamma}^{(1)} = 0$,  $\chi^{(0)} \in [0.3,0.4]$ and $\chi^{(1)} \in [0.4,0.55]$ (pairwise). The black line corresponds to the true values.
}
\label{fig:univsmulti}
\end{figure} 

\section{Analysis of heavy precipitation from the  CNRM model}\label{sec:appli}
Evidencing causality is more difficult for heavy rainfall than for extreme temperatures, because 
precipitation variability is greater in space and time and because extreme rainfall has heavier tails than temperatures (extreme rainfall often has $\gamma \approx 0.2$, see, e.g. \citet{katz02}). 
We work with simulated rainfall time series from the French global climate model of M\'et\'eo-France (CNRM) 
that belongs to the latest Coupled Model Intercomparison Project (CMIP6). We consider the winter months between the 1st of January 1985 until the 31st of August 2014 over the region defined by $-10$ to $40$ in longitude and $35$ to $60$ in latitude (corresponding to central Europe). Our factual and counterfactual worlds correspond to two historical runs, the second one of which has only natural forcings. We take the weekly maxima of  winter precipitation. As the number of years covers only three decades,  the rainfall series can be considered stationary in time within each world. 
Concerning their spatial structure, we apply the partitioning around medoids (PAM)  algorithm \citep{kaufman1990} to the counterfactual rainfall run. 
The difference with the original PAM version is that our ``distance'' between two locations $s$ and $t$  is tailored to threshold exceedances via   
\begin{equation*}
\widehat{d}_{st} =  \frac{1 - \widehat{\chi}_{st}}{2(3 -\widehat{\chi}_{st})},
\end{equation*}
where $\widehat{\chi}_{st}$ denotes the standard empirical estimator of the pairwise tail dependence coefficient \citep[see, e.g.][and references therein]{kiriliouk2016book}. Our approach is close to the one of \cite{bernard2013}, who focused on maxima instead of threshold exceedances. 
Figure \ref{fig:clusters.pdf} displays the spatial structure for $K=40$ clusters\footnote{Other values of $K$ were tested and provided similar patterns.}. 
Although no spatial coordinates were given to the algorithm, the clusters appear to be spatially homogeneous and climatologically coherent. As the multivariate GPD is tailored for asymptotic dependence, identifying dependent regions helps to improve its fit. In addition, such a spatial clustering makes the assumption of a constant shape parameter within a region more reasonable.
\begin{figure}[hb]
\centering
\subfloat{\includegraphics[width=1.1\textwidth]{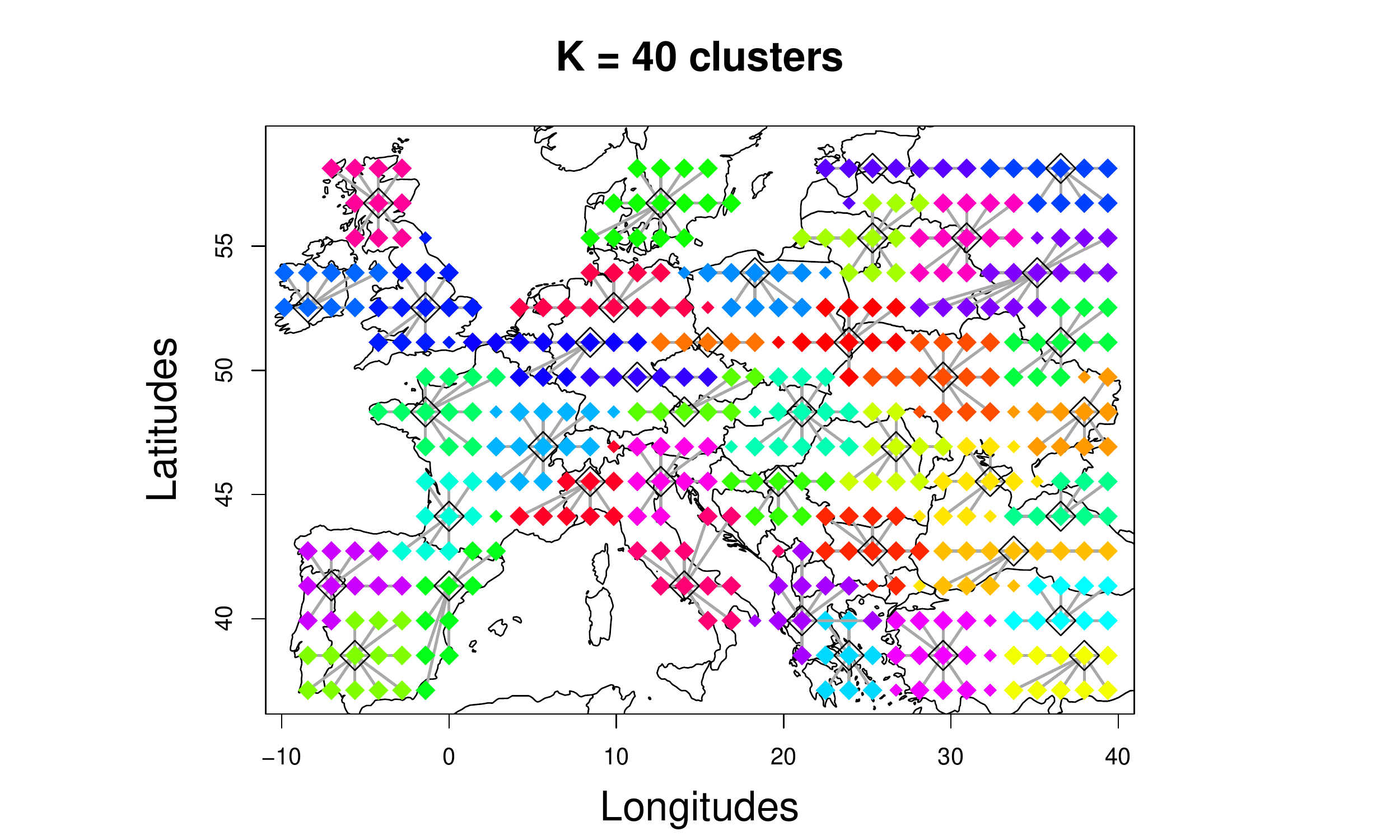}}
\caption{Clustering of weekly maximum winter precipitation in central Europe between January 1985 and August 2014, using the PAM algorithm with distance based on pairwise tail dependence coefficients.} 
\label{fig:clusters.pdf}
\end{figure}
We hence model each cluster independently, calculating the estimated multivariate PN based on $\widehat{p}_0, \widehat{p}_1$ defined in \eqref{eq:pest2}. 

Figures \ref{fig:PNworld5} and \ref{fig:PNworld50} show
necessity causation probabilities for the five-year and fifty-year return level respectively. The return levels were calculated based on quantiles of $\vc{w}^T X^{(0)}$ for each cluster, with equal weights. Both figures show the PN per cluster, calculated using equal weights (top) and optimal weights (bottom). 
The diameters of the black circles around the PN estimates are proportional to the length 
of bootstrap-based 95\%  confidence intervals. 
\begin{figure}[p]
\centering
\subfloat{\includegraphics[width=0.925\textwidth]{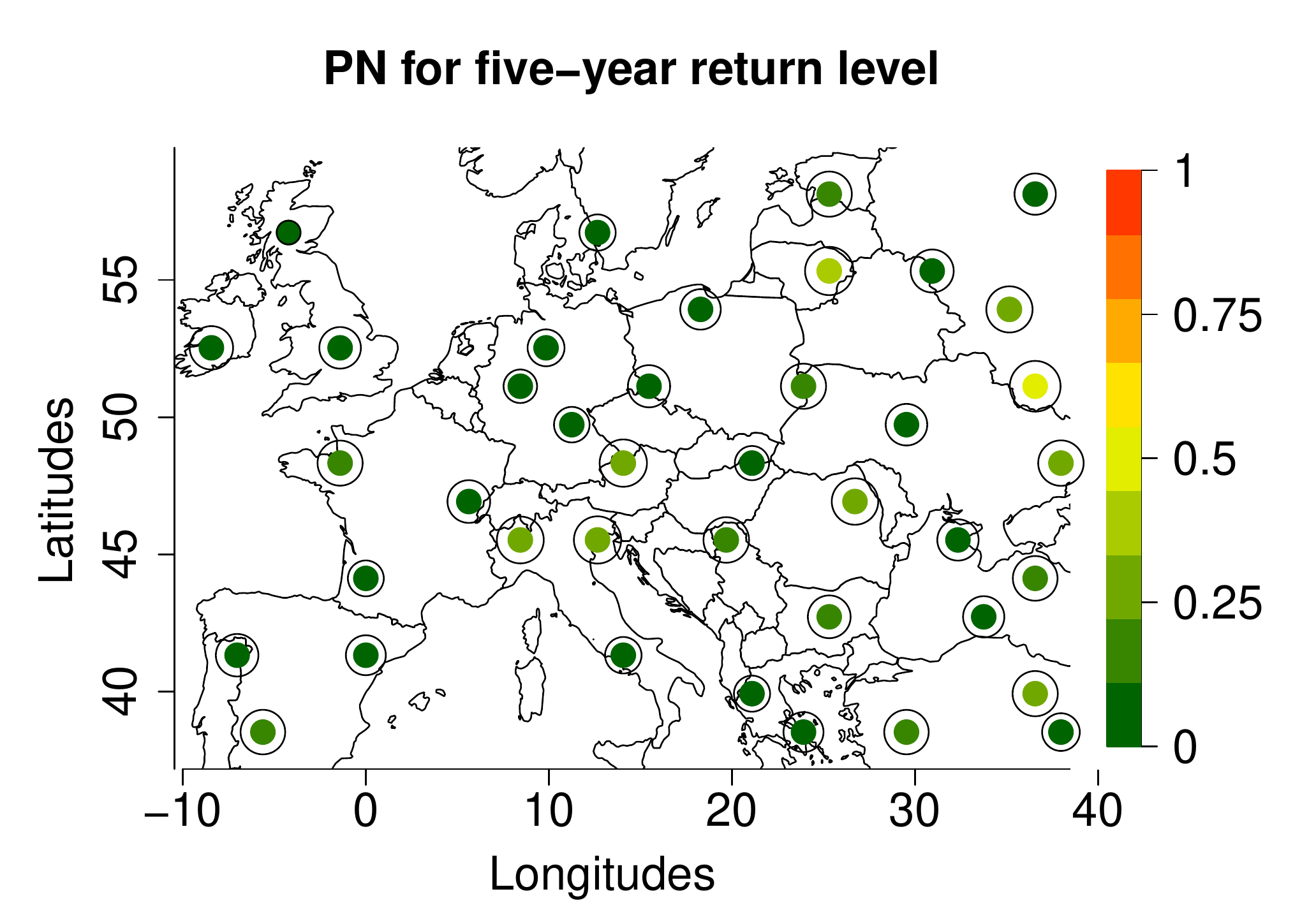}} \\
\vspace{-0.5cm}
\subfloat{\includegraphics[width=0.925\textwidth]{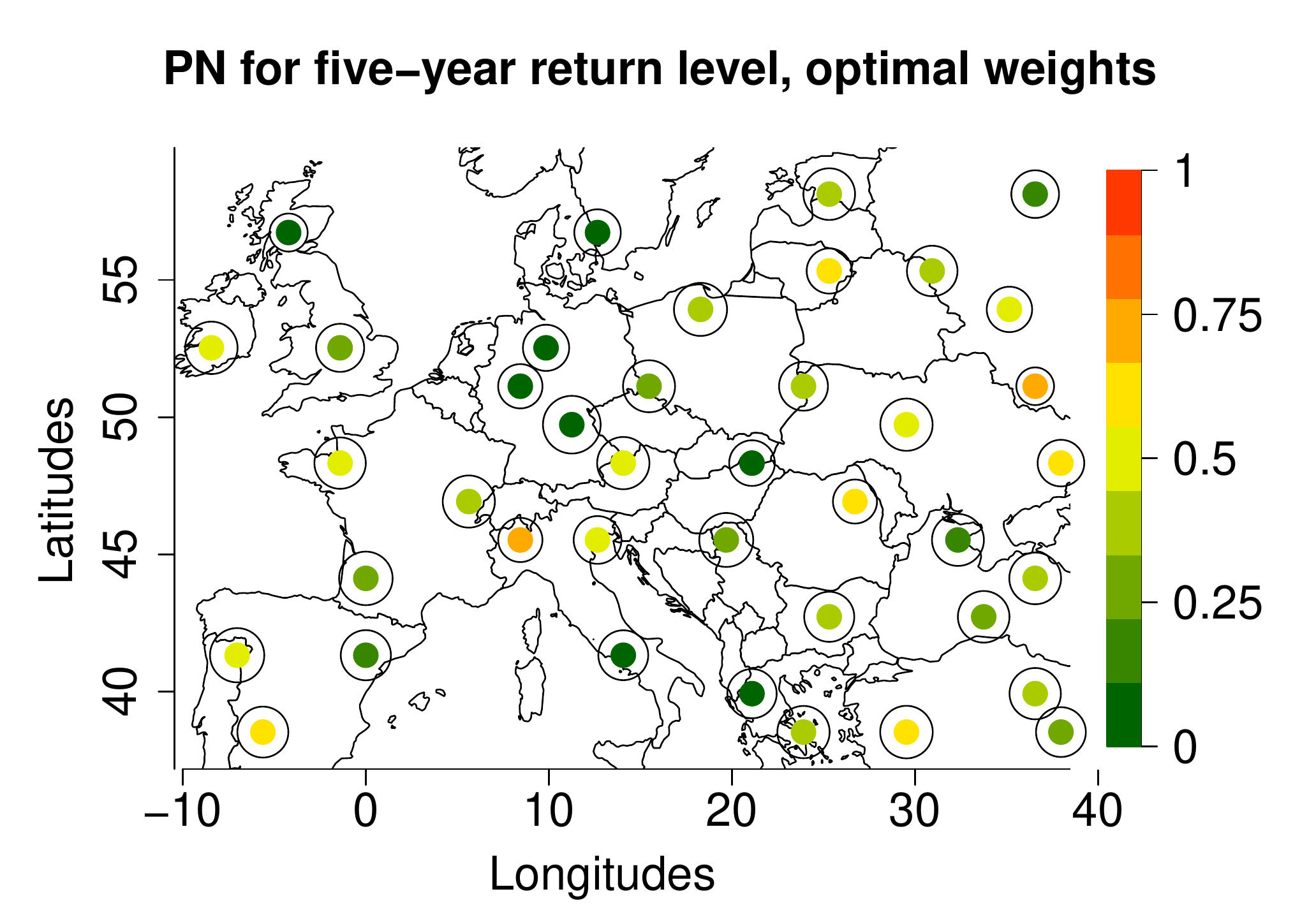}} 
\caption{Necessary causation probabilities for a five-year return level of weighted maximal weekly winter precipitation in the counterfactual world for each cluster, calculated using equal weights (top) and optimal weights (bottom). The diameters of the black circles around the estimates are proportional to the length 
of bootstrap-based 95\%  confidence intervals.  
} 
\label{fig:PNworld5}
\end{figure}
\begin{figure}[p]
\centering
\subfloat{\includegraphics[width=0.925\textwidth]{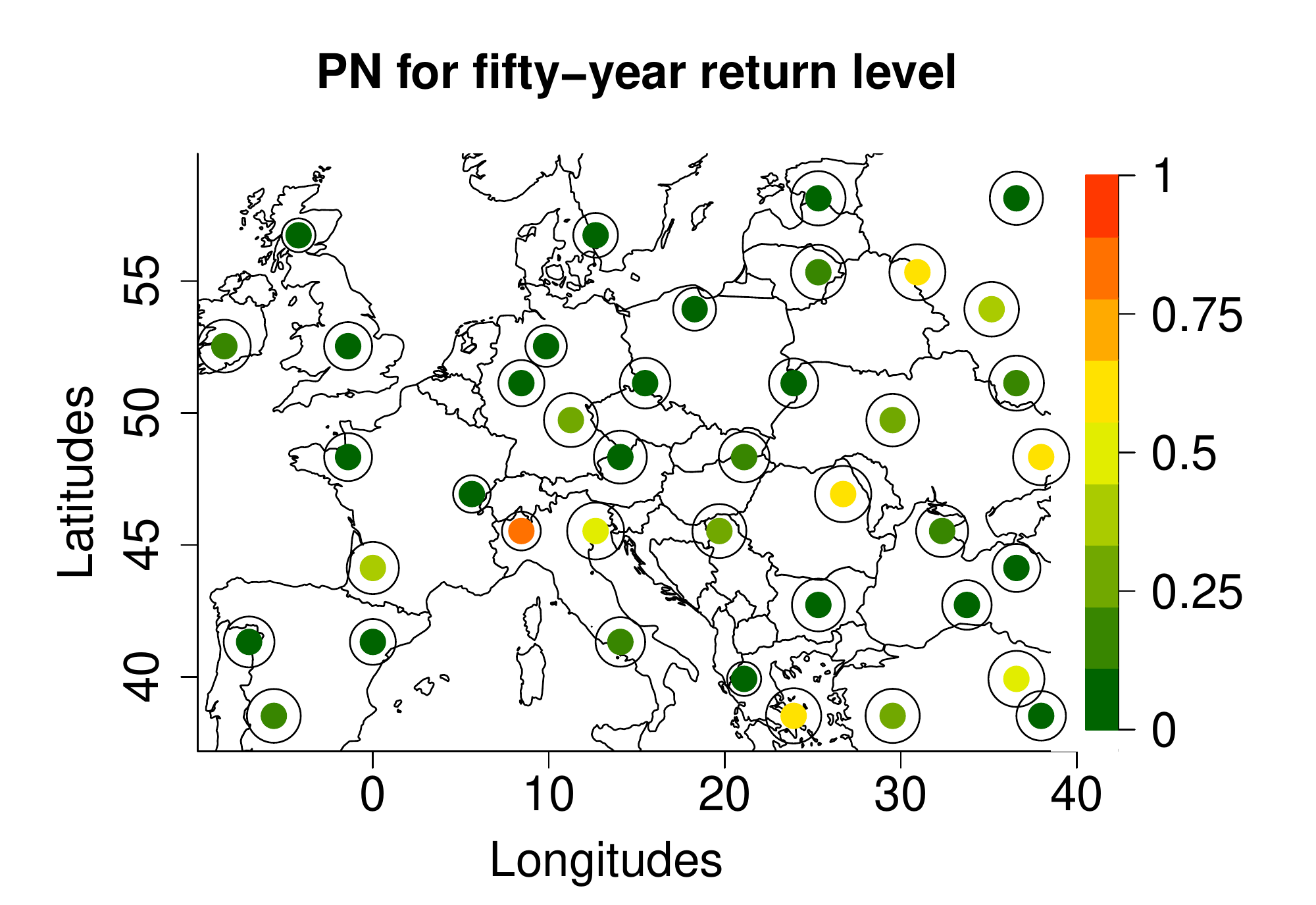}} \\
\vspace{-0.5cm}
\subfloat{\includegraphics[width=0.925\textwidth]{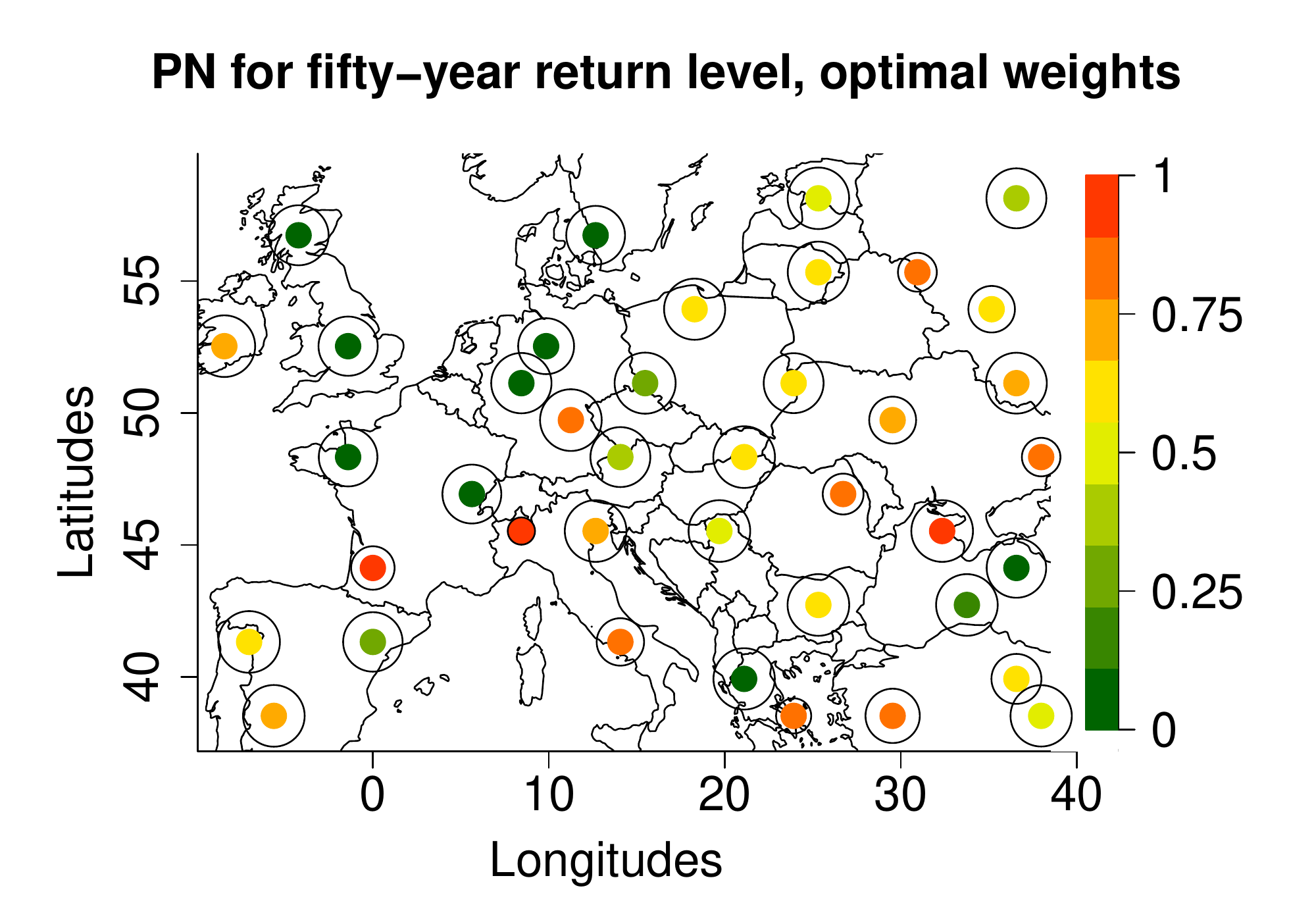}}
\caption{Necessary causation probabilities for a fifty-year return level of weighted maximal weekly winter precipitation in the counterfactual world for each cluster, calculated using equal weights (top) and optimal weights (bottom). The diameters of the black circles around the estimates are proportional to the length 
of bootstrap-based 95\%  confidence intervals.  
} 
\label{fig:PNworld50}
\end{figure}
Higher PN does not necessarily comes with higher uncertainty; see for instance the cluster around northern Italy, whose confidence interval is narrow for the five-year return level and even more so for the fifty-year return level.
Comparing the two panels of  Figure \ref{fig:PNworld5} shows that the differences between the factual and the counterfactual world are higher when using optimal weights. This feature is even more striking for the fifty-year return levels, see Figure \ref{fig:PNworld50}. 
Except for locations near the English channel, most points have a probability of necessary causation that is greater than $0.5$.
In particular, a few points like northern Italy shows a probability near one. 

In Section E of the supplementary material \citep{KirilioukNaveau20}, the above approach is compared to several univariate approaches for the five-year return level. We found that, even though patterns are similar (i.e., high PN around northern Italy), a univariate analysis leads to lower PN on average and to wider confidence intervals when PN is relatively high ($> 0.5$). Hence, a multivariate GPD approach enhances the causality message of a univariate analysis and aids in decreasing the uncertainty of the estimates. 

Finally, our analysis is not sufficient to conclude general climatological results about heavy rainfall. The patterns found here may be due to this specific climate model, internal climate variability or other sources of variability. An exhaustive analysis of all the CMIP6 models, in terms of computer resources and climatological expertise, is beyond the scope of this work.

\section{Discussion}\label{sec:discussion}
This paper illustrates that methods combining multivariate extreme-value theory and counterfactual theory could help climatologists working on causality and multivariate extremes \cite[see, e.g.][]{kim16,Zscheischler18}. An advantage of our approach is its simplicity: we consider an event to be extreme if the weighted average of a climatological random vector exceeds a high threshold, and EVT naturally suggests the multivariate GPD to model such multivariate threshold exceedances. While multivariate EVT models can take on complex parametric forms, the model we propose is easily estimated since linear projections of multivariate GPD vectors behave like a univariate GPD. When spatial dependence changes between the factual and the counterfactual world, a univariate analysis might under- or overestimate the causation probabilities, while the proposed approach will takes these changes into account. In addition, the application on heavy precipitation suggests that the multivariate approach can help in reducing the uncertainty of the estimates. Finally, the multivariate approach can highlight those grid points that maximise the probability of necessary causation through an adequate choice of the weights $\vc{w}$.

Some care is needed when applying the multivariate methodology. First of all, the model is restricted to homogeneous regions, since it assumes an equal shape parameter. This is a common assumption in multivariate EVT models. Moreover, the multivariate GPD is tailored for data that exhibit \emph{asymptotic dependence}, i.e., the extremes in each grid point are expected to occur together. Finally, the dataset under study  needs to be reasonably stationary in time. In future work, a non-stationary extension of the multivariate GPD could be proposed that would be highly relevant for longer periods of data from the factual world.

Another interesting research direction will be to extend the coupling between EVT and counterfactual theory to other types of extremes modeling in geosciences; see, e.g. 
\citet{Hammerling17} or \citet{Reich13} for a Bayesian hierarchical point of view, \citet{Shooter19} for asymptotic independence models, and \citet{Ragone18} for rare event algorithms.

\begin{supplement}[id=suppA]
 \stitle{Supplement to ``Climate extreme event attribution using 
 multivariate peaks-over-thresholds modeling  and counterfactual theory''}
  \slink[doi]{DOI}
 \sdatatype{.pdf}
  \sdescription{The supplement includes results on the Gaussian MGPD model, probability weighted moment inference, a simulation study and further analysis for the precipitation data.}
\end{supplement}

\bibliographystyle{imsart-nameyear} 
\bibliography{biblio}

\end{document}